\documentclass[11pt]{article}

\usepackage{jheppub}

\DeclareMathAlphabet\mathbfcal{OMS}{cmsy}{b}{n}
\usepackage{xfrac}
\usepackage{nicefrac}
\usepackage{empheq,fancybox}  
\usepackage{amsmath}	
\usepackage{bm}		    
\usepackage{bbm}
\usepackage{amssymb}	
\usepackage{makeidx}	
\usepackage{color} 		
\usepackage{graphicx} 	
\usepackage{enumitem}
\usepackage{multirow}	
\usepackage{times} 		
\usepackage{tensor} 	
\usepackage{slashed}
\usepackage[all]{hypcap}  
\usepackage[title]{appendix}
\usepackage{dsfont}
\usepackage{notoccite}
\usepackage{siunitx}

\newcommand{\xvec}{{\bm x}}

\newcommand{\kvec}{{\bm k}}

\newcommand{\RH}{\text{\sc rh}}
\newcommand{\HRH}{H_\RH}
\newcommand{\aRH}{a_\RH}

\newcommand{\TRH}{T_\RH}

\newcommand{\DM}{\text{\sc dm}}
\newcommand{\Scal}{\mathcal{S}}
\newcommand{\Rcal}{\mathcal{R}}
\newcommand{\eref}[1]{Eq.~(\ref{#1})}

\newcommand{\half}{{\textstyle\frac{1}{2}}}

\newcommand{\GeV}{\ \mathrm{GeV}}

\newcommand{\Mpl}{M_{\rm Pl}}

\usepackage{accents}

\newcommand{\RN}[1]{%
  \textup{\uppercase\expandafter{\romannumeral#1}}
}

\numberwithin{equation}{section}

\def\be{\begin{equation}}
\def\ee{\end{equation}}
\def\figs/B{B}
\def\bea{\begin{eqnarray}}
\def\eea{\end{eqnarray}}
\def\bg{\begin{eqnarray}}
\def\nd{\end{eqnarray}}
\def\log{{\rm log}}

\begin{document}

\title{Completely Dark Matter from Rapid-Turn Multifield Inflation}

\author[a]{Edward W.\ Kolb,}
\emailAdd{rocky.kolb@uchicago.edu}
\affiliation[a]{Kavli Institute for Cosmological Physics and Enrico Fermi Institute, University of Chicago, Chicago, IL 60637, USA}

\author[b]{Andrew J.\ Long,}
\emailAdd{andrewjlong@rice.edu}
\affiliation[b]{Department of Physics and Astronomy, Rice University, Houston, TX 77005, USA}

\author[c]{Evan McDonough,}
\emailAdd{e.mcdonough@uwinnipeg.ca}
\affiliation[c]{Department of Physics, University of Winnipeg, Winnipeg, MB, R3B 2E9, Canada}

\author[c,d]{and Guillaume Payeur}
\emailAdd{guillaume.payeur@mail.mcgill.ca}
\affiliation[d]{Department of Physics, McGill University, Montreal, QC, H3A 0G4 ,Canada}
\date{\today }

\abstract{
We study cosmological gravitational particle production as applied to ``rapid-turn'' models of inflation involving two scalar fields.  We are interested in the production of massive spin-0 particles that only interact gravitationally and provide a candidate for the dark matter. 
Specifically, we study two models of rapid-turn multifield inflation, motivated in part by the de Sitter swampland conjecture, that are distinguished by the curvature of field space and the presence or absence of field space `angular momentum' conservation.  We find that one of these models leads to insufficient particle production and cannot explain the observed dark matter relic abundance.  The second model is able to explain the origin of spin-0 dark matter via gravitational production, and we identify the relevant region of parameter space that is consistent with measurements of the dark-matter relic abundance, the dark-matter-photon isocurvature perturbations, and the spectrum of curvature perturbations that is probed by cosmological observations.  Our work demonstrates the compatibility of the de Sitter swampland conjecture with completely dark matter. }

\begingroup
\setlength{\parskip}{0.0cm}
\maketitle
\endgroup

\section{Introduction}\label{sec:intro}

The standard early-universe cosmology involves an early phase of accelerated expansion known as inflation.  It is usually assumed that the dynamics of inflation is driven by the potential energy $V(\phi)$ of a single scalar field $\phi(x)$, known as the \textit{inflaton} field.  In the quasi-de Sitter phase of inflation the cosmological expansion rate $H(t)$ must satisfy $\epsilon \equiv - \dot{H} / H^2 \ll 1$. 
During \textit{slow-roll} inflation this condition may be associated with a requirement on the flatness of the potential $\epsilon_V(\phi) \equiv (\Mpl^2 / 2) (V^\prime(\phi) / V)^2 \ll 1$.  Measurements of anisotropies in the cosmic microwave background (CMB) temperature and polarization are a powerful probe of the inflationary epoch.  Some once-favored models of inflation are now excluded; for example, Quadratic Inflation~\cite{Linde:1983gd} and Natural Inflation~\cite{Freese:1990rb} each predict a primordial spectrum of curvature perturbations that is in significant disagreement with measurements of the CMB furnished by the \textit{Planck} satellite telescope (2018)~\cite{Planck:2018jri} (see also \cite{Stein:2021uge}).  Other still-viable models of inflation have begun to receive renewed attention.  
Much of this recent work has explored models of inflation driven by the dynamics of multiple scalar fields, so-called \textit{multifield inflation} (see Refs.~\cite{Wands:2007bd,Gong:2016qmq} for reviews).  In the context of well-motivated ultraviolet (UV) embeddings, such as string theory, one generically expects multiple fields with comparable masses, whose collective dynamics govern the cosmological expansion during inflation. Moreover, such multifield dynamics can even `revive' an otherwise excluded model; for example, bringing Natural Inflation into agreement with \textit{Planck}~\cite{McDonough:2020gmn}. 

Theoretical considerations also provide valuable guidance when navigating the space of viable models of inflation.  One such consideration comes from the swampland program~\cite{Vafa:2005ui}, which seeks to connect low-energy effective field theories that provide a phenomenological description of Nature with high-energy theories that also accommodate the quantization of gravity (for a review, see~\cite{Brennan:2017rbf,Palti:2019pca}).  In particular the de Sitter (dS) swampland conjecture~\cite{Obied:2018sgi} asserts that effective field theories with a consistent UV completion in quantum gravity must satisfy the condition
\begin{equation}
	\label{eq:dsSwamp}
	V'(\phi) \gtrsim V(\phi) / \Mpl  \;,
\end{equation}
for every possible field configuration $\phi(x)$.  Here $V' \equiv \sqrt{G^{I J} \partial_I V \partial_J V}$ is the norm of the potential gradient defined using the metric $G^{IJ}$ on field space in the kinetic term of the scalar fields.  The dS swampland conjecture ostensibly forbids slow-roll inflation, which requires $\epsilon_V(\phi) < 1$, however it remains consistent with, and provides motivation for, multifield models of inflation, such as the ones we study here.  Moreover, the conjecture prevents $V(\phi)$ from having a minimum $(V'=0)$ at positive energy $(V>0)$ and thus, insofar as the cosmological constant is due to the potential energy of scalar fields, it forbids a cosmological constant as the observed dark-energy component of the universe.  One might naturally wonder if the swampland conjectures have anything to say about the other side of the dark universe, namely dark matter, and indeed many recent works have considered this possibility, e.g., Refs.~\cite{Agrawal:2019dlm,McDonough:2021pdg,Anchordoqui:2022txe,Gonzalo:2022jac}.

Planck data indicates the existence of dark matter to a statistical significance greater than 70$\sigma$~\cite{Planck:2018vyg}, and yet the identity and origin remain a mystery. Inflation provides a built-in minimal mechanism to produce the observed dark-matter abundance: So-called ``Gravitational Particle Production'' (GPP)~\cite{Chung:1998zb,Chung:1998ua,Chung:1998rq,Kolb:1998ki,Kuzmin:1998uv, Chung:2001cb,Chung:2004nh} requiring only a particle coupled to gravity. As interest in non-thermal dark candidates has risen, so too has interest in inflation as a particle factory. GPP has been studied for a variety of particle spins (such as spin-$1/2$~\cite{Chung:2011ck,Ema:2019yrd,Herring:2020cah,Maleknejad:2022gyf,Hashiba:2022bzi}, spin-$1$~\cite{Graham:2015rva,Ema:2019yrd,Ahmed:2019mjo,Ahmed:2020fhc,Kolb:2020fwh}, spin-$3/2$~\cite{Kallosh:1999jj,Giudice:1999yt,Giudice:1999am,Kallosh:2000ve,BasteroGil:2000je}, spin-$2$~\cite{Albornoz:2017yup}, and spin-$s$ ($s>2$) fields~\cite{Alexander:2020gmv}), and in inflation models consistent with current data (see e.g.,~\cite{Ling:2021zlj,Garcia:2022vwm}). 

Recently a class of inflation models was developed that does not require a flat potential. So-called ``rapid-turn'' inflation models utilize angular momentum to resist the downward pull of a steep potential gradient.  Models include hyperinflation \cite{Brown:2017osf,Bjorkmo:2019aev}, angular inflation \cite{Christodoulidis:2018qdw}, sidetracked inflation \cite{Garcia-Saenz:2018ifx}, and orbital inflation \cite{Achucarro:2019pux}, among others \cite{Aragam:2020uqi,Aragam:2019omo}; see  Ref.~\cite{Bjorkmo:2019fls} for a review.  The UV completion of these models has been recently discussed in Ref.~\cite{Aragam:2021scu}. These models, of which Refs.~\cite{Brown:2017osf,Christodoulidis:2018qdw, Garcia-Saenz:2018ifx} predate the de Sitter swampland conjecture, are interesting in their own right as new inflation models consistent with Planck and testable by next-generation experiments.  They are also naturally consistent with the dS swampland conjecture.  
 
Rapid-turn inflation models may be classified by their field-theory construction, notably whether the field space manifold is hyperbolic (as in Refs.~\cite{Brown:2017osf,Bjorkmo:2019aev,Christodoulidis:2018qdw,Garcia-Saenz:2018ifx}) or flat \cite{Achucarro:2019pux}, and whether the potential energy $V$ respects (\cite{Brown:2017osf}) or breaks  (\cite{Achucarro:2019pux}) the isometries of the field-space manifold (see \cite{Achucarro:2019mea} for related work). Models may alternatively be classified by their dynamics, and in particular, in the context of two-field models with a radial field $\phi$ and an angular field $\theta$,  the evolution of the field-space angular momentum $J$:
\begin{itemize}
\item  {\bf Hyperbolic Inflation:} Models with a hyperbolic field space naturally feature an exponential stretching at large radii (responsible for such phenomena as $\alpha$-attractor inflation \cite{Kallosh:2013yoa}), leading to an exponential enhancement of the angular momentum $J =L^2\dot{\theta}\sinh^2(\phi /L)$, where $L$ is the curvature radius of the hyperbolic space. If the potential energy respects the isotropy of the field space, then the comoving angular momentum is conserved, ${\rm d}(a^3 J)/{\rm d}t=0$, and hence $J \propto a^{-3}$ redshifts exponentially fast during inflation, but may nonetheless play an important role in the dynamics. We refer to this class of models as hyperbolic inflation.
\item {\bf Monodromy Inflation:} A second class of models forgoes the exponential enhancement of $J$,  taking instead a flat field space. These models break angular momentum conservation via an explicit angular dependence of the potential energy, or monodromy, in which case angular momentum satisfies ${\rm d}(a^3 J)/{\rm d}t=a^3{\rm d}V/{\rm d}\theta$. With an appropriate an appropriate choice of $V(\phi,\theta)$, these models can realize a near-constant $J$, leading to slow-roll inflation. We refer to this class of models as monodromy inflation.
\end{itemize}

In this work we study the gravitational production of dark matter in rapid-turn inflation models that are consistent with the dS swampland conjecture. We  focus on the two representative models of Refs.\ \cite{Brown:2017osf} and  \cite{Achucarro:2019pux}, which implement the inflationary dynamics using a hyperbolic inflation model and a monodromy inflation model, respectively.  For both of these models, we perform a detailed numerical analysis of GPP for a scalar spectator field, including the dependence on the scalar's mass and nonminimal coupling to gravity.  We find that rapid-turn inflation in hyperbolic inflation~\cite{Brown:2017osf} cannot produce sufficient particles to explain the observed relic abundance of dark matter.  This conclusion is reversed for rapid-turn inflation in monodromy inflation~\cite{Achucarro:2019pux}, for which GPP is qualitatively similar to the well-studied case of quadratic inflation. The origin of this difference lies in the parameter fine-tunings needed for perturbations of the inflationary fields to match the Planck constraints on $n_s$ and $A_s$: Rapid-turn inflation for hyperbolic inflation is characterized by an instability of isocurvature perturbations, sourcing a growing curvature perturbation,\footnote{While the model presented in Ref.~\cite{Christodoulidis:2018qdw} is an exception to this, evidence from supergravity~\cite{Aragam:2021scu} suggests this is a rather generic feature.} and as a result, the fit to Planck amplitude of scalar perturbations $A_s$ forces upon these models an exponentially lower energy scale of inflation than would be naively anticipated, a feature that is absent in the monodromy inflation case. The relatively low scale of inflation for rapid-turn inflation in hyperbolic inflation prevents the model from producing any sizeable amount of dark matter.

Looking beyond the scope of this work, we observe that the dS swampland conjecture may be consistent with a gravitational origin of dark matter.  Thus, while the dS swampland conjecture forbids the simplest (and an arguably `boring') dark-energy candidate, namely a cosmological constant, it is not in conflict with the simplest (and an arguably `boring') dark-matter candidate, namely dark matter with only gravitational interactions. We note that gravitationally produced dark matter exhibits a rich phenomenology (see, e.g., \cite{Carney:2022gse}), making it an exciting avenue for future research.

The remainder of this article is structured as follows.  In Sec.~\ref{sec:inflation} we introduce the two models of rapid-turn multi-field inflation that we study.  In Sec.~\ref{sec:gpp} we first provide a short review of GPP, and then we present the results of our numerical study, which take the form of comoving number density spectra of gravitationally produced particles.  In Sec.~\ref{sec:dm} we assess whether GPP provides a viable explanation for the origin of dark matter, and we illustrate the various observational constraints in the parameter space of our model.  Finally we summarize and discuss our results in Sec.~\ref{sec:discussion}.  The article is extended by four appendices:  App.~\ref{app:hyper} contains an analytical analysis of GPP; App.~\ref{app:isocurvature} contains additional details relating to the isocurvature calculation; App.~\ref{app:relic abundance} contains a derivation of the relic abundance in terms of comoving number density; and App.~\ref{app:relic quadratic} contains a brief discussion to contrast our results with the familiar model of quadratic inflation. 

\textit{Conventions.}  We denote the reduced Planck mass by $\Mpl = (8 \pi G_N)^{-1/2} = 2.435 \times 10^{18} \GeV$.  We adopt the Landau-Lifshitz timelike conventions for the signature of the metric: ($-,+,+$)  in the introductory material in Misner, Thorne, and Wheeler \cite{MTW}.

\section{Rapid-Turn Multifield Inflation}
\label{sec:inflation}

The conventional wisdom is that inflation with $\epsilon = - \dot{H}/H \ll 1$ may be achieved by taking a sufficiently flat potential for the inflaton field, i.e. $\epsilon_V = (\Mpl^2/2) (V^\prime/V)^2 \ll 1$.  Indeed, quadratic inflation with $V(\phi) \propto \phi^2$ implies $\epsilon = \epsilon_V$ during the slow-roll phase. However, rapid-turn multifield inflation provides an exception to conventional wisdom. As emphasized in Ref.~\cite{Achucarro:2018vey}, theories with more than one inflaton field admit an alternative class of dynamics: in a two-field model with one `radial' field and one `angular' field, inflation with $\epsilon \ll 1$ can occur even for $\epsilon_V \gtrsim 1$. This is made possible by an angular momentum in field space, and the corresponding evolution of the fields in the direction orthogonal to the gradient of the potential.  As stated in the introduction, there are by now many model realizations of this idea \cite{Christodoulidis:2018qdw,Garcia-Saenz:2018ifx, Achucarro:2019pux, Brown:2017osf, Bjorkmo:2019aev,Aragam:2020uqi,Aragam:2019omo}.  In the remainder of this section, we summarize two models of rapid-turn multifield inflation that are consistent with the dS Swampland conjecture. 

\subsection{Class 1: Hyperbolic Inflation}
\label{subsec:hyperinflation}

Multifield inflation in hyperbolic inflation was proposed in Refs.~\cite{Brown:2017osf}, \cite{Christodoulidis:2018qdw}, and \cite{Garcia-Saenz:2018ifx}, going by the names hyperinflation, angular inflation, and sidetracked inflation, respectively. (See also  Refs.~\cite{Bjorkmo:2019aev,Aragam:2020uqi,Aragam:2019omo}.)  A generic feature of these models is that the isocurvature perturbations can have a negative mass-squared, i.e., can be tachyonic, due to the curvature of the field space manifold, a feature first emphasized in the context of ``geometric destabilization'' \cite{Renaux-Petel:2015mga}. Indeed, hyperinflation and sidetracked inflation are characterized by isocurvature perturbations that grow outside the horizon, seeding a growing curvature perturbation. Angular inflation differs from these by a careful cancellation between  positive and negative contributions, leading to stable isocurvature perturbations at all times during the angular inflation phase, however the supergravity analysis of  Ref.~\cite{Aragam:2021scu} suggests that tachyonic isocurvature is a generic feature of rapid-turn models. Guided by this, in what follows we focus on the simple case of hyperinflation \cite{Brown:2017osf,Bjorkmo:2019aev}, which is itself a special case of sidetracked inflation (see Ref.~\cite{Bjorkmo:2019fls}).  

We study a specific realization of hyperinflation \cite{Brown:2017osf} with two fields $\varphi^1 \equiv \phi$ and $\varphi^2 \equiv \theta$ whose dynamics are governed by an action of the form  
\begin{equation}\label{eq:hyperinflation_action}
	S = \int \! \mathrm{d}^4 x \, \sqrt{-g} \, \biggl[ \frac{1}{2} G_{IJ}(\phi) \partial_\mu \varphi^I \partial^\mu \varphi^J - V(\phi) \biggr] 
	\;.
\end{equation}
We take the field space metric to be 
\begin{align}
	G_{IJ}(\phi) = \begin{pmatrix}
	1 & 0\\
	0 & L^2 \sinh^2(\phi/L) 
	\;
\end{pmatrix},
\end{align}
which corresponds to a hyperbolic space $\mathbb{H}^2$ where $\phi$ and $\theta$ are the respective `radial' and `angular' coordinates. 
The field space is characterized by a scalar curvature $R_{\rm field-space} = -2/L^2$, which is controlled by the parameter $L$.  

The hyperinflation background dynamics are relatively insensitive to the choice of potential. For concreteness, we take the scalar potential to be 
\begin{equation}
\label{eq:V}
	V(\phi)  = V_0 \tanh^2\Big(\frac{\phi - v}{f}\Big) \, e^{\lambda \phi/\Mpl}.
\end{equation}
This toy model is chosen for simplicity of the dynamics and observables.  
During inflation the inflaton field amplitude is large, $|\phi-v|\gg f$, and the potential is nearly exponential, $V \sim V_0 \, e^{\lambda \phi/ \Mpl}$.  
Rapid-turn inflation on an exponential potential has been studied previously~\cite{Mizuno:2017idt} , and it is known to lead to constant-roll evolution of the Hubble parameter ${\rm d}\epsilon/{\rm d}N =0$, allowing for simple analytic expressions for the spectral index and other observables.  After inflation, the inflaton field amplitude settles to a local minimum at $\phi = v$, and since $V(v) = 0$, the dS swampland conjecture~\eqref{eq:dsSwamp} is still satisfied.  The inflationary observables are insensitive to $v$ and $f$ since since $\phi \sim \Mpl  \gg v,f$ during inflation.  

In an Friedmann-Lema\^{i}tre-Robertson-Walker (FLRW) universe, $(\mathrm{d} s)^2 = (\mathrm{d} t)^2 - a(t)^2 |\mathrm{d} {\bm x}|^2$, the action~\eqref{eq:hyperinflation_action} for the homogeneous part of the inflaton fields is governed by 
\begin{align}
    S=\int \! \mathrm{d}^4x \, a^3 \bigg[ \frac{1}{2}\dot{\phi}^2+\frac{1}{2}L^2\sinh^2\Big(\frac{\phi}{L}\Big)\dot{\theta}^2-V(\phi) \bigg] 
    \;.
\end{align}
The corresponding equations of motion are written as 
\begin{equation}\label{eq:hyperphi}
\begin{split}
	& \ddot{\phi} + 3 H \dot{\phi} - \frac{L}{2} \sinh\Big(\frac{2\phi}{L}\Big) \, \dot{\theta}^2 = - V^\prime(\phi) \\ 
	& \text{and} \quad 
	\frac{\rm d}{{\rm d}t} \left( a^3 J \right) = 0
	\qquad \text{where} \qquad 
	J \equiv \sinh^2\Big(\frac{\phi}{L}\Big) \, \dot{\theta}     \;,
\end{split}
\end{equation}
The radial field $\phi(t)$ obeys a Klein-Gordon equation, while the angular field $\theta(t)$ obeys an equation expressing the conservation of field-space angular momentum $J$.  The Friedman equation is
\begin{equation}
\label{eq:hyperH}
    3 \Mpl^2 H^2 =  \frac{1}{2}\dot{\phi}^2+\frac{1}{2}L^2\sinh^2\Big(\frac{\phi}{L}\Big)\, \dot{\theta}^2+V(\phi) 
\end{equation}
where the energy density of both fields contribute.  

This model exhibits inflation $(\epsilon \equiv -\dot{H}/H^2 \ll 1)$ even for a steep potential, $V'/V \gtrsim \Mpl $. For a potential satisfying the steepness condition
 \begin{align}
    \frac{V^\prime(\phi)}{3H} > 3HL,
\end{align}
the model has a slow-roll ($\epsilon\ll 1$) inflation solution given by
\begin{equation}
\begin{split}
\label{eq:hyperinflation-attractor}
    \dot{\phi} & =-3HL \\
    L\sinh\Big(\frac{\phi}{L}\Big) \, \dot{\theta} &=\sqrt{L\partial_\phi V-(3HL)^2} 
    \;.
\end{split}
\end{equation}
This solution is a local attractor in field space \cite{Bjorkmo:2019fls}.  
From  Eq.~\eqref{eq:hyperinflation-attractor}, one may infer the duration of slow-roll inflation in terms of the number of e-folds $N_{\rm tot}$ defined by ${\rm d}N = H {\rm d}t$, as $N_{\rm tot} \sim \phi_i / 3L$ where $\phi_i$ is $\phi$ at the beginning of inflation, which must be at least $\phi_i \gtrsim 180 L$.  It follows that the initial angular momentum is exponentially enhanced $J_i = \sinh^2(\phi_i/L) \, \dot{\theta}_i \sim e^{6 N_\mathrm{tot}} \, \dot{\theta_i}$, and it continues to have an important impact on the dynamics despite the exponential redshifting during expansion, $J(t) \propto a(t)^{-3} \approx e^{- 3 H t}$. 

As a fiducial example, we consider the model specified by the parameters
\begin{equation}
\label{eq:paramshyper}
    \lambda=2 \,\, ,\,\, L=0.0054 \Mpl \,\, , \,\,  V_0= 2.6 \times 10^{-36} \Mpl^4 \,\, ,  \,\, v=f=10^{-2}\Mpl 
    \;.
\end{equation}
These parameters are chosen to match the measured CMB observables, discussed further below.  

\begin{figure}[t]
    \centering
    \includegraphics[width=\textwidth]{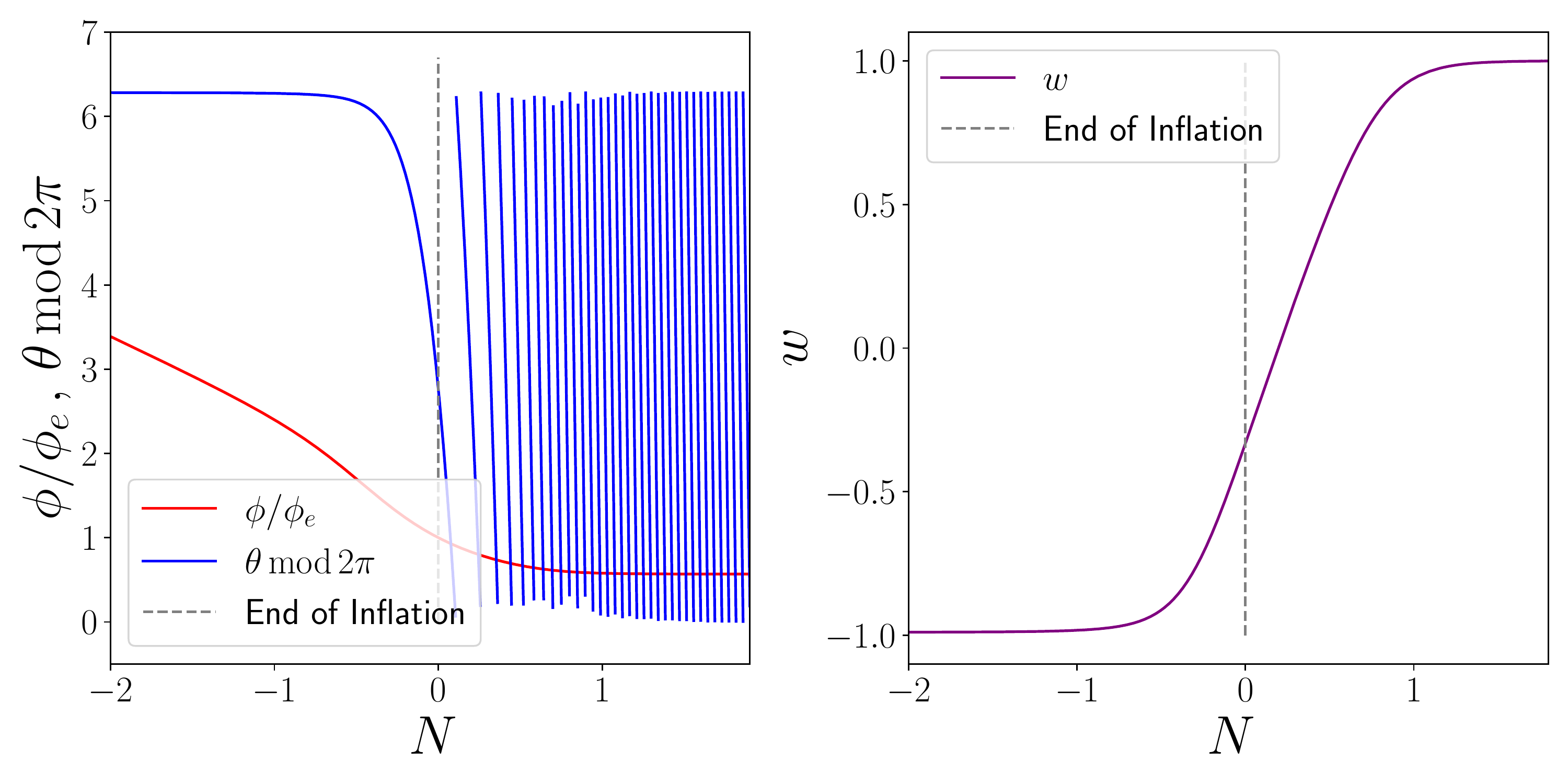}
    \caption{\label{fig:hyperinflation-background}
Evolution of the homogeneous background fields $\phi(t)$ and $\theta(t)$ for a fiducial model of hyperbolic inflation.  Time is parametrized by the number $N$ of $e$-folds in the scale factor, normalized to $N=0$ at the end of inflation.  
\textit{Left:}  The field amplitudes are shown, normalized to their values $\phi_e$ and $\theta_e$ at the end of inflation.  The radial field variable $\phi$ decreases monotonically throughout inflation, becoming fixed to $\phi = v$ after inflation; whereas the angular field variable $\theta$ grows monotonically, completing many $2\pi$ cycles.  The evolution is consistent with the conservation of angular momentum $a^3 J \approx e^{3N} e^{2 \phi/L} \dot{\theta}$.  
\textit{Right:}  The evolution of the equation of state $w(t)$ reveals a transition from a quasi-de Sitter phase of inflation ($w \approx -1$) to a post-inflationary phase of kination ($w \approx 1$).  }
\end{figure}

We numerically solve Eqs.~\eqref{eq:hyperphi} and \eqref{eq:hyperH} to find the evolution of the homogeneous field backgrounds $\phi(t)$, $\theta(t)$, and $a(t)$.  The initial conditions are chosen to be $\phi_i = 180L$ and $\theta_i = 0$, which ensures $50$ $e$-foldings of inflation, and the initial field velocities are set by Eq. (\ref{eq:hyperinflation-attractor}). We present our results in Fig.~\ref{fig:hyperinflation-background}, which shows the evolution of the fields near the end of inflation $N = 0$.  We also show the evolution of the equation of state parameter $w\equiv p/\rho$, with $p$ and $\rho$ given by
\begin{equation}
\begin{split}
    \rho & = \frac{1}{2}\dot{\phi}^2+\frac{1}{2}L^2\sinh^2\Big(\frac{\phi}{L}\Big) \, \dot{\theta}^2+V(\phi) \\ 
    p & =\frac{1}{2}\dot{\phi}^2+\frac{1}{2}L^2\sinh^2\Big(\frac{\phi}{L}\Big) \, \dot{\theta}^2-V(\phi) 
    \;.
\end{split}
\end{equation}
During inflation the $V(\phi)$ terms dominate and $w \approx -1$ corresponding to the quasi-de Sitter phase.  
After inflation the $\dot{\theta}^2$ terms dominate, since $\dot{\theta}$ must grow to maintain angular momentum conservation, and $w \approx 1$ corresponds to a phase of kination (kinetic energy domination).  
This is a hallmark feature of hyperbolic inflation, which distinguishes it from models of single-field slow-roll inflation that typically have $w \approx 0$ after inflation, while the inflaton oscillates about a quadratic minimum of its potential.  

Inflationary observables are derived from the spectrum of perturbations in the inflaton fields.  Perturbations in multifield inflation are discussed in detail in Ref.~\cite{Amin:2014eta}, while perturbations in rapid-turn multifield models are discussed in Ref.~\cite{Bjorkmo:2019fls}.
In this class of models, the evolution of perturbations is governed by an effective mass matrix that depends on both derivatives of the potential and on the Riemann tensor on field space (see Ref.~\cite{Amin:2014eta}). Due to the hyperbolic field space, and its negative curvature $R_{\rm field-space} = - 2 / L^2$, the effective mass-squared matrix admits a tachyonic direction (negative eigenvalue).  
The tachyonic mode grows throughout inflation, which exponentially enhances the curvature perturbations.  
To ensure that the observed amplitude of scalar perturbations is predicted, a commensurate reduction in the inflationary Hubble scale is required.  

Using the software package PyTransport \cite{Mulryne:2016mzv}, we numerically solve for the evolution of perturbations during inflation and derive predictions for inflationary observables.  Our results for the observables reproduce that given in \cite{Mizuno:2017idt} for the case of an exponential potential. The model specified by Eq.~\eqref{eq:paramshyper} yields an amplitude $A_s = 2.1 \times 10^{-9}$ and spectral tilt $n_s = 0.97$ of the scalar power spectrum that is consistent with the measured values~\cite{Planck:2018jri}.  
The tensor-to-scalar ratio is predicted to be exponentially small, $r\sim (V_0/\Mpl^4)/A_s \sim 10^{-25}$, putting a primordial B-mode signal out of reach.  The non-Gaussianity was argued in Ref.~\cite{Brown:2017osf} to satisfy the usual consistency relation $f_{\rm nl} \propto (n_s-1)$; though interesting phenomenology is possible \cite{Fumagalli:2019noh}. 

In particular, note that a particularly low energy scale of inflation is needed in order to reproduce the measured $A_s$.  
We take $V_0^{1/4} \sim 10^{-9} \Mpl \sim 10^{8} \GeV$ corresponding to a Hubble scale at the end of inflation of $H_e \sim 1 \GeV$, or a maximum possible reheat temperature of $T_\RH^{\rm max} \sim \sqrt{H_e\Mpl} \sim 10^9 \GeV$.  
A small $V_0$ is required because the scalar perturbations grow exponentially in models of hyperbolic inflation, and an over-production of unwanted curvature perturbations is avoided by lowering the inflationary scale by a commensurate factor.  
The small $V_0$ is notable, since classic models of large-field inflation with Planck-scale excursions for the inflaton field typically provide a much larger $V_0^{1/4} \sim 10^{16} \Mpl$ and $H \sim 10^{14} \GeV$.  
Moreover, studies of gravitational dark matter production during inflation typically assume $H \gg \mathrm{GeV}$, and we will see that hyperbolic inflation's low inflationary Hubble scale leads to a strong suppression of gravitational particle production. 

Finally, we note the consistency of this model with the de Sitter Swampland conjecture.  
For the model parameters in Eq.~\eqref{eq:paramshyper}, we calculate $V'/V = 2/\Mpl$ at large values of the inflaton field $\phi$.  
Consequently, the model is consistent with the dS swampland conjecture~\eqref{eq:dsSwamp}.

\subsection{Class 2: Monodromy Inflation}
\label{subsec:flat field-space inflation}

As a second class of rapid-turn multifield inflation, we consider the models studied in Refs.~\cite{Achucarro:2012yr,Achucarro:2019pux}.   Unlike hyperbolic inflation, the field-space manifold is flat and the isocurvature perturbations are stable. Akin to the ``monodromy'' model of axion inflation \cite{Silverstein:2008sg,McAllister:2008hb}, the scalar potential carries an explicit dependence on the angular field variable, which induces an angular momentum.  However, unlike hyperbolic inflation, the angular momentum remains approximately constant during slow-roll inflation.

We consider a two-field model with `radial' field $\phi(x)$ and `angular' field $\theta(x)$, whose dynamics are governed by the action
\begin{equation}
	S = \int \! \mathrm{d}^4x \, \sqrt{-g} \, \biggl[ \frac{1}{2} (\partial \phi)^2 + \frac{1}{2}\phi^2 (\partial \theta)^2 - V(\theta,\phi) \biggr] 
	\;.
\end{equation}
In Ref.~\cite{Achucarro:2012yr} the scalar potential is taken to be $V_0 - \alpha \theta + (m^2/2) (\phi - \phi_0)^2$.  The linear dependence on $\theta$ is reminiscent of the original proposal of axion monodromy inflation \cite{Silverstein:2008sg,McAllister:2008hb}.  The parameter $\alpha$ can be positive or negative; the sign of $\alpha$ merely sets the sign of $\dot{\theta}$ during slow-roll inflation.  Since this potential is unbounded from below in the $\theta$ direction, but we seek to study post-inflationary dynamics, we instead consider the potential,
\begin{equation}
\label{eq:Vflat}
	V(\theta, \phi) = V_0 \, e^{ - \alpha (\theta-\theta_i)/V_0} \tanh^2 \left(\theta/f\right)+ \frac{1}{2}m^2 (\phi - \phi_0)^2
	\;,
\end{equation}
which has been modified to introduce a minimum for $\theta$.  After inflation both fields $\phi$ and $\theta$ reach a local quadratic minimum of the scalar potential, which is a rather generic feature of scalar potentials.  

Assuming an FLRW spacetime, we derive the equations of motion for the homogeneous background fields $\phi(t)$ and $\theta(t)$, as well as the scale factor $a(t)$; these equations are found to be 
\begin{equation}\label{algebraic-equations}
\begin{split}
	\ddot{\phi} + 3H\dot{\phi}-\phi\dot{\theta}^2+m^2(\phi-\phi_0) & = 0 \, , \\
	\ddot{\theta} + 3 H \dot \theta  +\frac{2}{\phi}\dot{\phi}\dot{\theta} + \frac{1}{\phi^2}\frac{\partial V}{\partial \theta} & = 0 \, , \\
3 \Mpl^2 H^2 & =  \frac{1}{2}\dot{\phi}^2 + \frac{1 }{2}  \phi^2 \dot \theta^2 + V(\theta,\phi) \, .
\end{split}
\end{equation}
During slow-roll inflation, the solutions of these equations are well approximated by 
\begin{equation}\label{aprox-background-sol}
	\phi = \sqrt{\frac{|\alpha| \Mpl}{\sqrt{3 V_0}  m}} 
	\qquad \text{and} \qquad 
	\dot \theta = - m 
	\ , 
\end{equation}
which results in $J=\phi^2\dot{\theta}\sim\mathrm{const}.$  This solution branch corresponds to $\alpha < 0$, and there is another branch with $\alpha > 0$ and $\dot{\theta} = +m$.  This describes circular motion in field space at a constant radius and angular frequency. 

\begin{figure}[t]
	\centering
	\includegraphics[width=\textwidth]{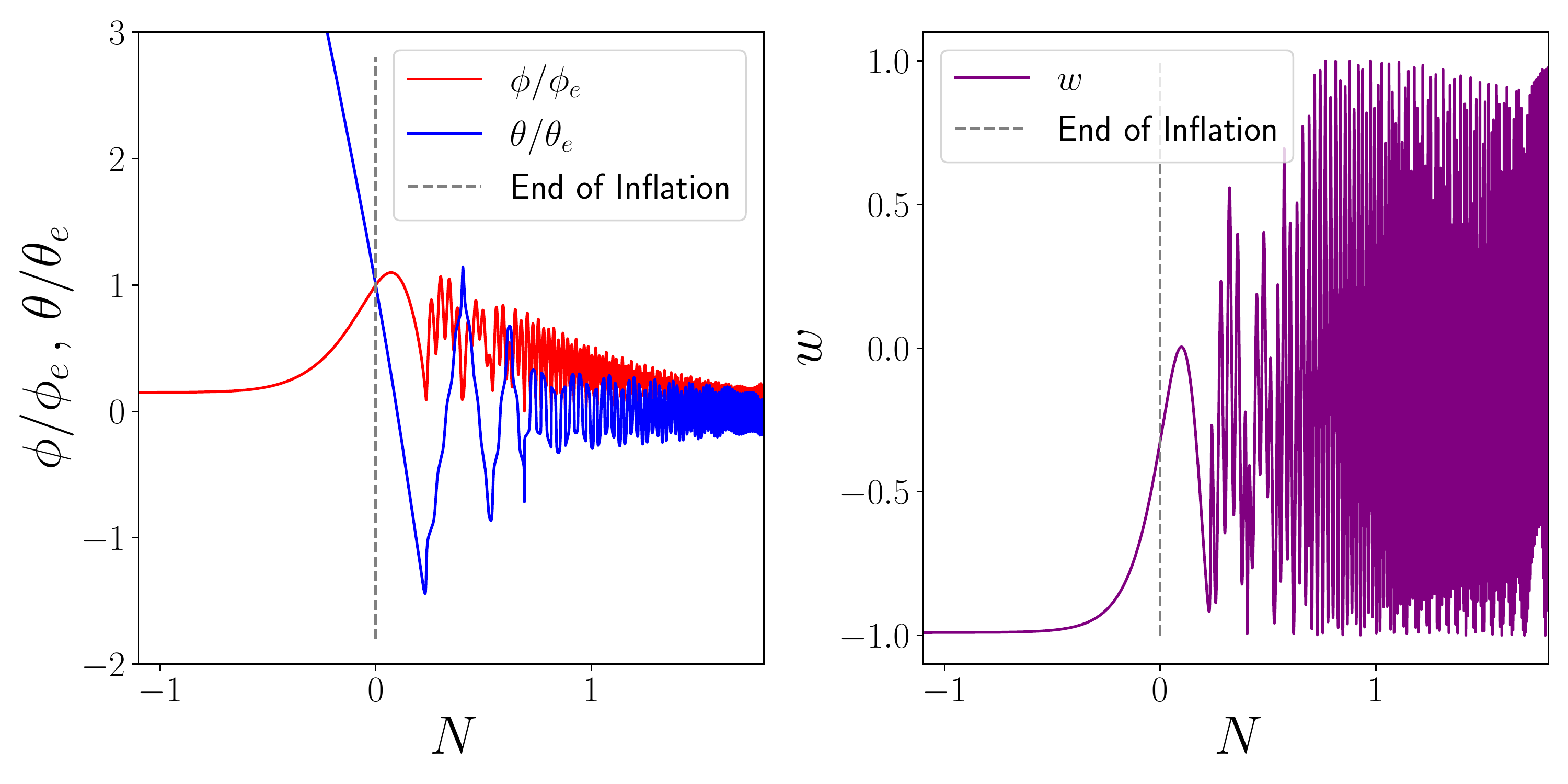}
	\caption{\label{fig:phi-and-theta-flat-space}
Evolution of the homogeneous background fields $\phi(t)$ and $\theta(t)$ for a fiducial model of flat-field-space inflation.  The notation here is the same as in Fig.~\ref{fig:hyperinflation-background}.  Inflation ends when $\theta$ crosses the minimum of its potential, triggering a recoil in $\phi$ followed by oscillations about the minimum. The equation of state following inflation has time-averaged value $\langle w \rangle = 0$, corresponding to matter domination.  }
\end{figure}

As a fiducial example, we consider the model with parameters given by 
\begin{eqnarray}
\label{eq:paramsflat}
    V_0=7.49\times10^{-10} \Mpl^4 \,\, ,\,\,
    \theta_i=2000 \,\, ,\,\,
    \alpha=-5.05\times10^{-13} \Mpl^4, \nonumber\\
    f=10\,\, ,\,\,
    \phi_0=6.80\times10^{-4} \Mpl \,\, , \,\, m=4.00 \times 10^{-4} \Mpl 
    \;.
\end{eqnarray}
We solve the equations of motion numerically to study the evolution of the homogeneous background fields $\phi(t)$, $\theta(t)$, and scale factor $a(t)$.  The initial conditions are given by Eq.~\eqref{aprox-background-sol}.   We present our numerical results in Fig.~\ref{fig:phi-and-theta-flat-space}, which shows the evolution of the field amplitudes and effective equation of state near the end of inflation.  During inflation $\phi$ remains approximately constant while $\theta$ decreases.  At the end of inflation, $\phi$ receives a `kick' and subsequently both fields oscillate rapidly.  The effective equation of state oscillates between $w = -1$ and $1$, such that it time averages to $\langle w \rangle = 0$, and the FLRW cosmology is effectively matter dominated.   

We calculate the inflationary observables numerically and validate these results with the analytic expressions provided in Ref.~\cite{Achucarro:2012yr}.  Unlike hyperbolic inflation, this model does not exhibit a tachyonic instability for the isocurvature perturbations.   The fiducial parameters in Eq.~\eqref{eq:paramsflat} lead to $n_s = 0.97$ and $A_s=2.1\times 10^{-9}$, which are consistent with cosmological observations.  Mapping on to an effective single-field model with sound speed $c_s$ via the expression $P_\Rcal = H^2/(8\pi^2 \epsilon c_s)$ \cite{Achucarro:2012yr} we find the squared sound speed is $c_s^2 = 0.04$. Ref.~\cite{Achucarro:2012yr} provides an expression for the tensor-to-scalar ratio $r$, which evaluates to $r \simeq 16 \epsilon c_s \simeq 0.02$, and which is below the Planck+BICEP/KECK upper bound $r<0.044$ \cite{Tristram:2020wbi}, while remaining within reach of future experiments~\cite{CMB-S4:2016ple}.  The non-Gaussianity is peaked in the equilateral configuration, and Ref.~\cite{Achucarro:2012yr} finds the relation $f_{nl} ^{\rm eq} \sim -80 \epsilon^2/r^2 \sim -10$, which is well within the Planck constraint $f_{nl} ^{\rm eq} = -26 \pm 47$ \cite{Planck:2019kim} while remaining in reach of near-term future experiments~\cite{CMB-S4:2016ple}.  The Hubble parameter at the end of inflation is $H_e \sim 5.8 \times 10^{-6} \Mpl \approx 1.4 \times 10^{13} \GeV$, which sets a maximum reheat temperature of $T_\RH ^{\rm max} \sim 10^{16} \GeV$. 

Finally, we note the compatibility of this model with the de Sitter swampland conjecture.  During inflation the gradient of the scalar potential is $\Mpl^{-1} |V'/V| \sim m \phi/ V_0 \sim 2$, which is $\mathcal{O}(1)$ and therefore consistent with the dS swampland conjecture \eqref{eq:dsSwamp}.

\section{Gravitational Particle Production}\label{sec:gpp}

Gravitational particle production (GPP) is a mechanism by which particles are produced via their interactions with gravity. 
Particularly, during cosmic inflation, quantum fields can be excited as a result of the expansion of the universe, which at late times may be interpreted as particle production. GPP has been studied in great detail, beginning historically with Refs.~\cite{Parker:1969au,Parker:1974qw, Fulling:1974zr,Ford:1986sy, Anderson:1987yt,Lyth:1996yj} and more recently in e.g., Refs.~\cite{Kolb:2021xfn,Hashiba:2021npn,Basso:2022tpd}. While GPP is an effect of tiny magnitude, it is possible for the energy density of particles produced gravitationally to be large at late times in comparison to other fields. In this case, these particles may constitute the observed dark matter. 
Since the original works \cite{Chung:1998zb,Chung:1998ua,Chung:1998rq,Kolb:1998ki,Kuzmin:1998uv, Chung:2001cb,Chung:2004nh}, many such dark matter candidates have been proposed \cite{Chung:2011ck,Kallosh:1999jj,Giudice:1999yt,Giudice:1999am,Graham:2015rva,Ema:2019yrd,Ahmed:2019mjo,Kallosh:2000ve,BasteroGil:2000je,Albornoz:2017yup,Kolb:2020fwh,Ahmed:2020fhc,Herring:2020cah,Alexander:2020gmv,Ling:2021zlj,Maleknejad:2022gyf,Hashiba:2022bzi,Garcia:2022vwm}. 

In what follows we provide a short review of the framework of gravitational particle production, focusing on a scalar spectator field.

\subsection{Lightning Review}

Consider a real scalar field $\psi(x)$ with mass $m_\chi$ that interacts only gravitationally.  
The dynamics of this field and the metric $g_{\mu\nu}(x)$ are governed by the action 
\begin{equation}
	S[\psi(x),g_{\mu\nu}(x)]=\int \! \mathrm{d}^4x \sqrt{-g} \bigg[ \frac{1}{2} \Mpl^2 R+ \frac{1}{2} g^{\mu\nu} \partial_\mu \psi \partial_\nu \psi - \frac{1}{2} m^2 \psi^2 + \frac{1}{2} \xi R \psi^2 \biggr] 
	\;.
\end{equation}
The dimensionless parameter $\xi$ parametrizes the scalar field's coupling to gravity~\cite{Chernikov:1968zm,Callan:1970ze,Bunch:1980br,Bunch:1980bs,Birrell:1982ix,Odintsov:1990mt,Buchbinder:1992rb,Parker:2009uva,Markkanen:2013nwa}, and $\xi \neq 0$ is considered a `nonminimal' coupling.  Even if $\xi$ vanishes at tree-level, a non-vanishing value is generically induced by loop corrections \cite{Callan:1970ze,Bunch:1980br,Bunch:1980bs,Birrell:1982ix, Odintsov:1990mt, Buchbinder:1992rb,Faraoni:2000gx,Parker:2009uva,Markkanen:2013nwa,Kaiser:2015usz}.  Moreover, the nonminimal coupling is a dimension-4 operator consistent with the symmetries of the action, and thus should be included from the perspective of effective field theory. 

Specializing to a spatially-flat FLRW spacetime $\mathrm{d} s^2 = a(\eta)^2 [ \mathrm{d} \eta^2 - |\mathrm{d} {\bm x}|^2]$ with conformal time $\eta$, the action is expressed as 
\begin{align}
	S[\psi(\eta,\xvec)] & = \int \! \mathrm{d}\eta \, \mathrm{d}^3\!\xvec \, \left[ \half a^2 \left( \partial_\eta \psi \right)^2 - \half a^2 \left(\nabla \psi \right)^2 - \half a^4 m^2 \psi^2 + \half a^4 \xi R \psi^2 \right] 
	\;,
\end{align}
The kinetic term is canonically normalized in terms of a comoving field variable $\chi(\eta,\xvec) = a(\eta) \psi(\eta,\xvec)$.  As a quantum operator, the comoving field is expressed as a Fourier expansion 
\begin{align}
	\chi(\eta,\xvec) = \int \! \! \frac{\mathrm{d}^3\kvec}{(2\pi)^3} \Big(\hat{a}_\kvec \, \chi_\kvec(\eta) \, e^{i\kvec\cdot\xvec} + \hat{a}_\kvec^\dagger \, \chi_\kvec^*(\eta) \, e^{-i\kvec\cdot\xvec}\Big)
	\;,
\end{align}
where modes are labeled by a comoving wavevector $\kvec$.  The annihilation and creation operators, $\hat{a}_\kvec$ and $\hat{a}_\kvec^\dagger$, satisfy the canonical commutation relations.  The mode function $\chi_\kvec(\eta)$ for wavevector $\kvec$ is only a function of the comoving wavenumber $k = |\kvec|$, and it solves the equation of motion 
\begin{align}
\label{eq:EOMchi}
	\partial_\eta^2 \chi_k(\eta) + \omega_k^2(\eta) \, \chi_k(\eta) = 0
	\;,
\end{align}
where the comoving squared angular frequency is 
\begin{align}\label{eq:omegaksq}
    \omega_k^2 = k^2 + a^2(\eta) \Big( m^2 + \frac{1}{6} (1-6\xi) R(\eta) \Big)
    \;,
\end{align}
and $R(\eta)$ the Ricci scalar.  We are interested in the solution that is the positive-frequency mode at asymptotically early time, which corresponds to the Bunch-Davies vacuum initial conditions: 
\begin{equation}
\begin{split}
	\chi_k(\eta) & \xrightarrow[\eta\to-\infty]{}\sqrt{\frac{1}{2k}}e^{-ik\eta},\\
	\partial_\eta\chi_k(\eta) & \xrightarrow[\eta\to-\infty]{} -i\sqrt{\frac{k}{2}}e^{-ik\eta}
	\;.
\end{split}
\end{equation}
The amount of particle production is related to the amplitude of the negative-frequency mode at late time, which is proportional to the coefficient $\beta_k$ of the Bogoliubov transformation that links the two bases~\cite{Birrell:1982ix}.  One can calculate $|\beta_k|$ by solving Eq.~\eqref{eq:EOMchi} and evaluating 
\begin{align}
\label{eq:bogoluibov}
    |\beta_k|^2 = \frac{\omega_k}{2}|\chi_k|^2+\frac{|\partial_\eta \chi_k|^2}{2\omega_k}-\frac{1}{2}.
\end{align}
In the late-time limit ($\eta\to+\infty$), the comoving particle number density is given by
\begin{align}
\label{eq:nk}
	n_\chi a^3 	= \frac{1}{a_e^3H_e^3} \int \! {\rm d}\log \, k \ n_k
    	\qquad \text{where} \qquad 
	n_k \equiv \frac{k^3}{2\pi^2}|\beta_k|^2 
	\;.
\end{align}
The spectrum of produced particles $n_k$ gives the comoving number density of particles with comoving wavenumber $k$.
The contribution to GPP from inflaton dynamics after inflation can also be described as a scattering process~\cite{Ema:2015dka,Ema:2018ucl,Chung:2018ayg,Kaneta:2022gug}; i.e., inflaton field oscillations in a quadratic potential correspond to a condensate of cold inflaton particles that may annihilate via gravity to a dark sector.  

A quick diagnostic probe of particle production is given in terms of the dimensionless adiabaticity parameter $A_k$, which we define as,
\begin{align}
\label{eq:adiabaticity parameter}
    A_k = \frac{\partial_\eta \omega_k }{\omega_k ^2}.
\end{align}
Adiabatic evolution, under which a positive-frequency mode at early times evolves to a positive-frequency mode at late times without any mixing of positive and negative frequency modes, occurs provided that $A_k$ is small at all times:
\begin{equation}
    \left|A_k\right| \ll 1 \,\,\,\,\,\, ({\rm Adiabatic \, \, evolution} ) .
\end{equation}
The efficiency of gravitational particle production is enhanced when adiabaticity is more strongly violated, i.e., if $|A_k|\gtrsim 1$ at some intermediate time. 

\subsection{GPP in Rapid-Turn Inflation Models}

In what follows, we numerically solve Eq.~\eqref{eq:EOMchi} for the evolution of the mode functions, construct the Bogoliubov coefficients $\beta_k$ using Eq.~\eqref{eq:bogoluibov}, and compute the spectrum $n_k$ using Eq.~\eqref{eq:nk}. 

\begin{figure}[t]
	\centering
	\includegraphics[width=0.5\textwidth]{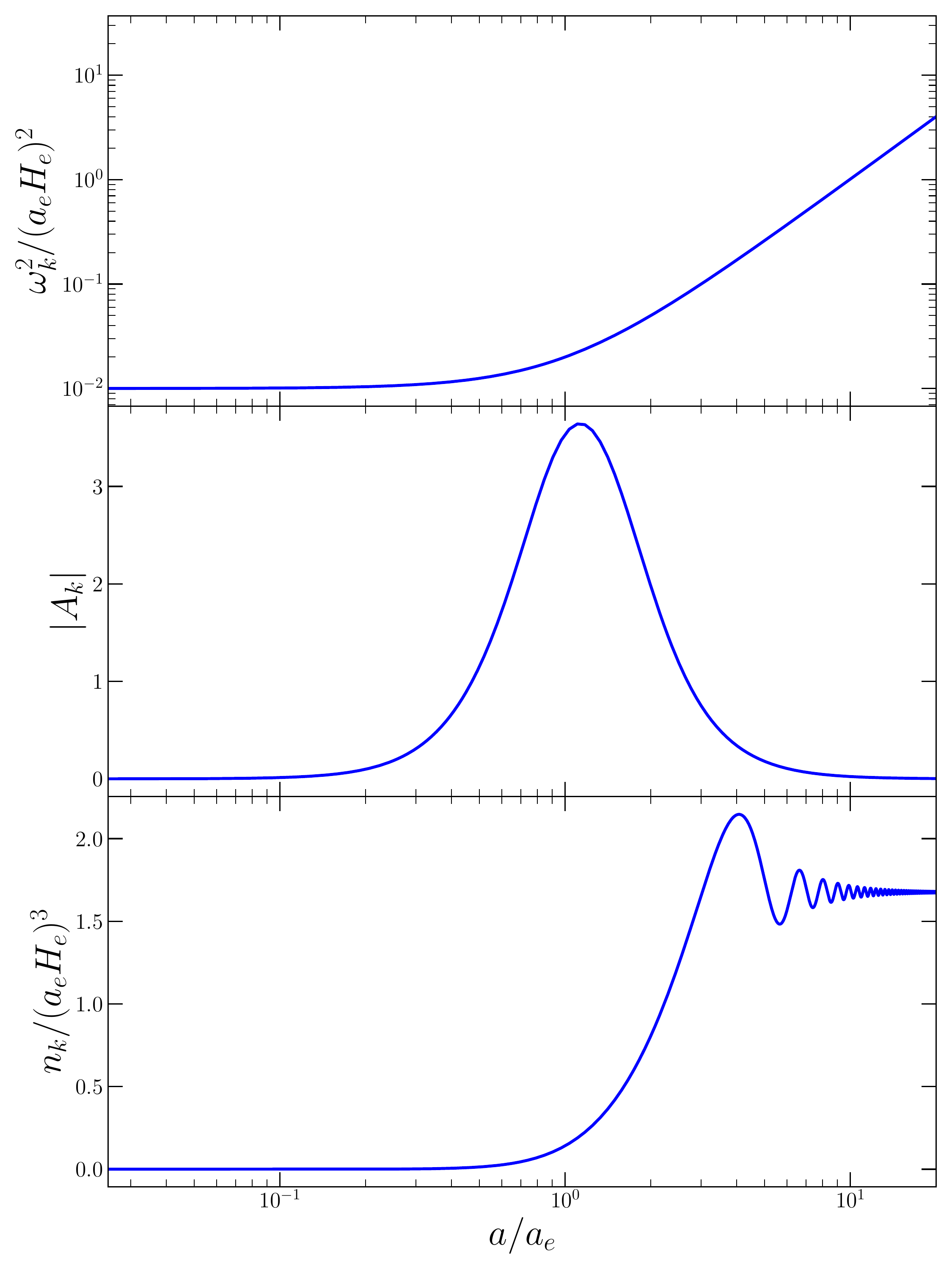}
	\caption{\label{fig:adiabaticity}
An illustration of non-adiabatic mode evolution leading to gravitational particle production.  For this example we take an FLRW background driven by hyperbolic inflation with Hubble parameter $H_e \simeq 1.8 \GeV$ and scale factor $a = a_e$ at the end of inflation, we take the scalar spectator mass to be $m_\chi = 0.1 H_e$, we take $\xi = 1/6$ corresponding to a conformal coupling with gravity, and we take a Fourier mode with comoving wavenumber $k = 0.1 a_e H_e$.  We show the evolution, near the end of inflation, of the comoving squared angular frequency $\omega_k^2$, the adiabaticity parameter $A_k = \partial_\eta \omega_k / \omega_k$, and the comoving number density spectrum $n_k$.  For this example, the `sudden' change in $\omega_k$ at the end of inflation $a/a_e = 1$ induces a large non-adiabaticity $|A_k| > 1$ and an associated particle production $n_k \neq 0$.  
}
\end{figure}

To build intuition for these results, we consider an illustrative numerical example showing the interplay of adiabaticity and particle production. 
We consider GPP in hyperbolic inflation with a scalar spectator having mass $m_\chi = 0.1 H_e$, comoving wavenumber $k = 0.1 a_e H_e$, and conformal coupling $\xi=1/6$.  We show $\omega_k^2$ in Fig.~\ref{fig:adiabaticity} along with the adiabaticity parameter $A_{k}(\eta)$ and the comoving number density $n_{k}(\eta)$ of gravitationally produced particles.  The dispersion relation \eqref{eq:omegaksq} evolves from $\omega_k ^2 \approx k^2$ at very early times to $\omega_k ^2 \approx a^2 m^2$ at late times.  The transition between these two regimes corresponds to a spike in the adiabaticity parameter $A_{k}(\eta)$, which in turn leads to a sharp rise in $n_{k}(\eta)$, followed by damped oscillations to an asymptotic late-time value, which determines the particle number of that $k$-mode. 

\begin{figure}[t]
	\centering
	\includegraphics[width=0.49\textwidth]{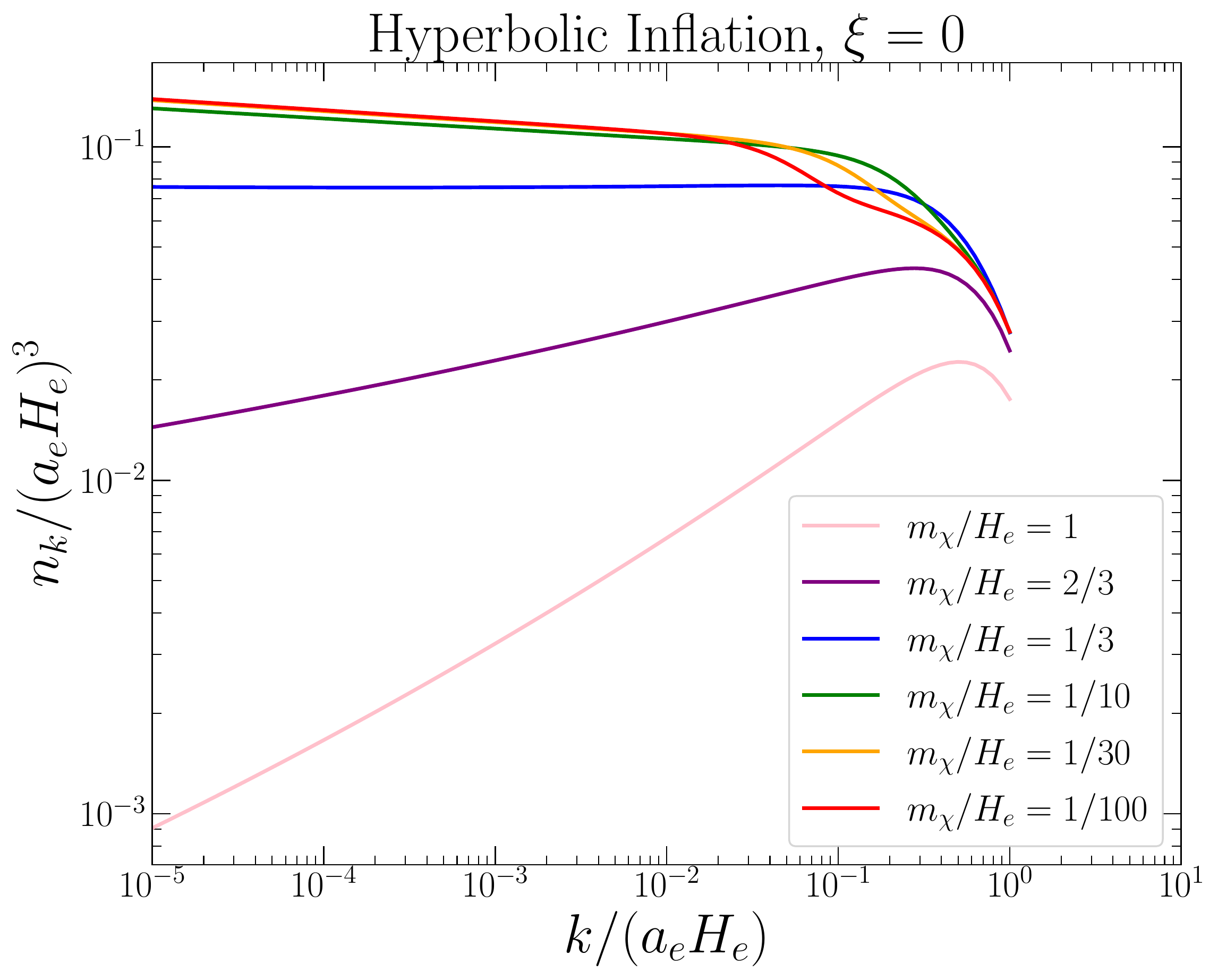}
	\includegraphics[width=0.49\textwidth]{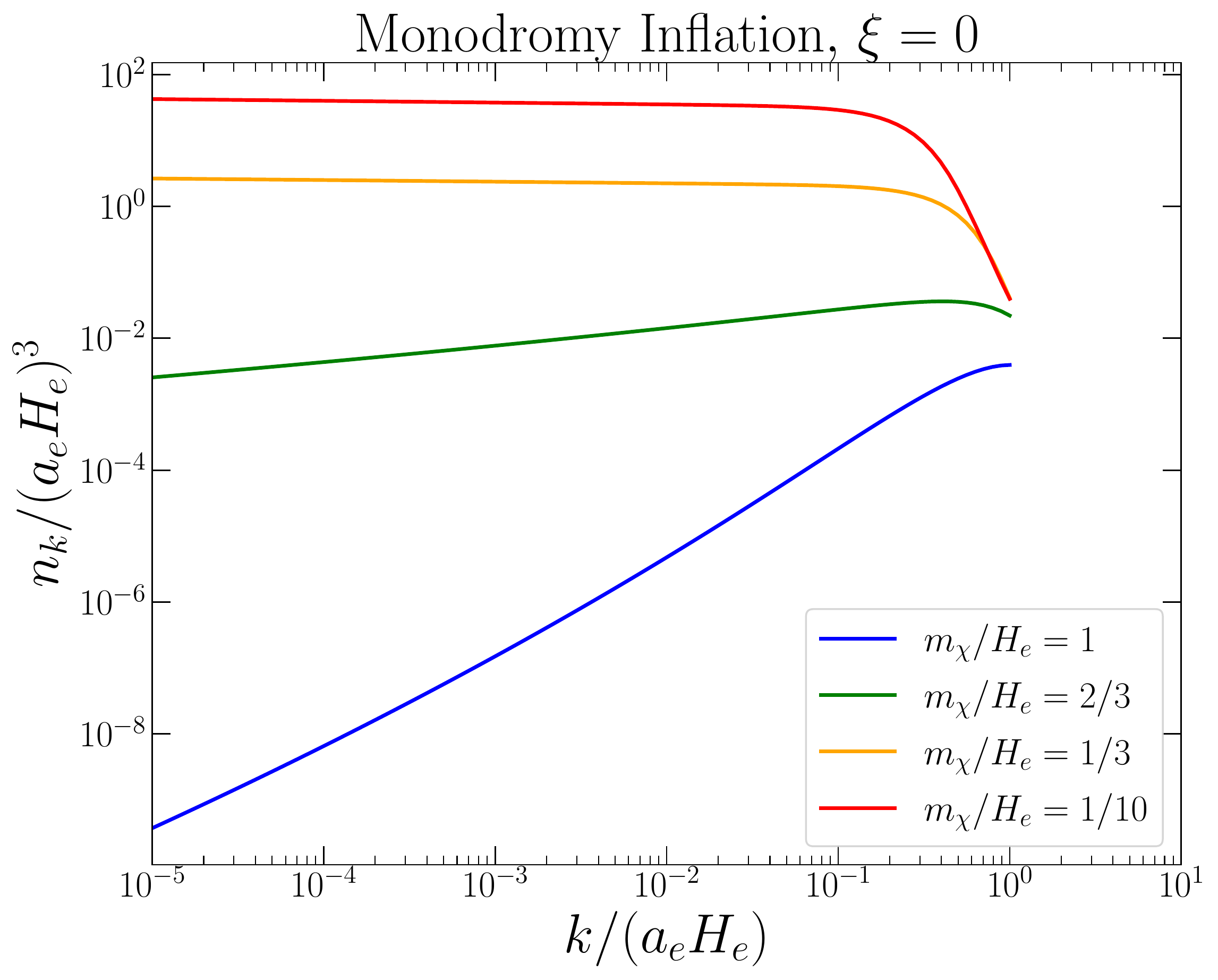}
	\includegraphics[width=0.49\textwidth]{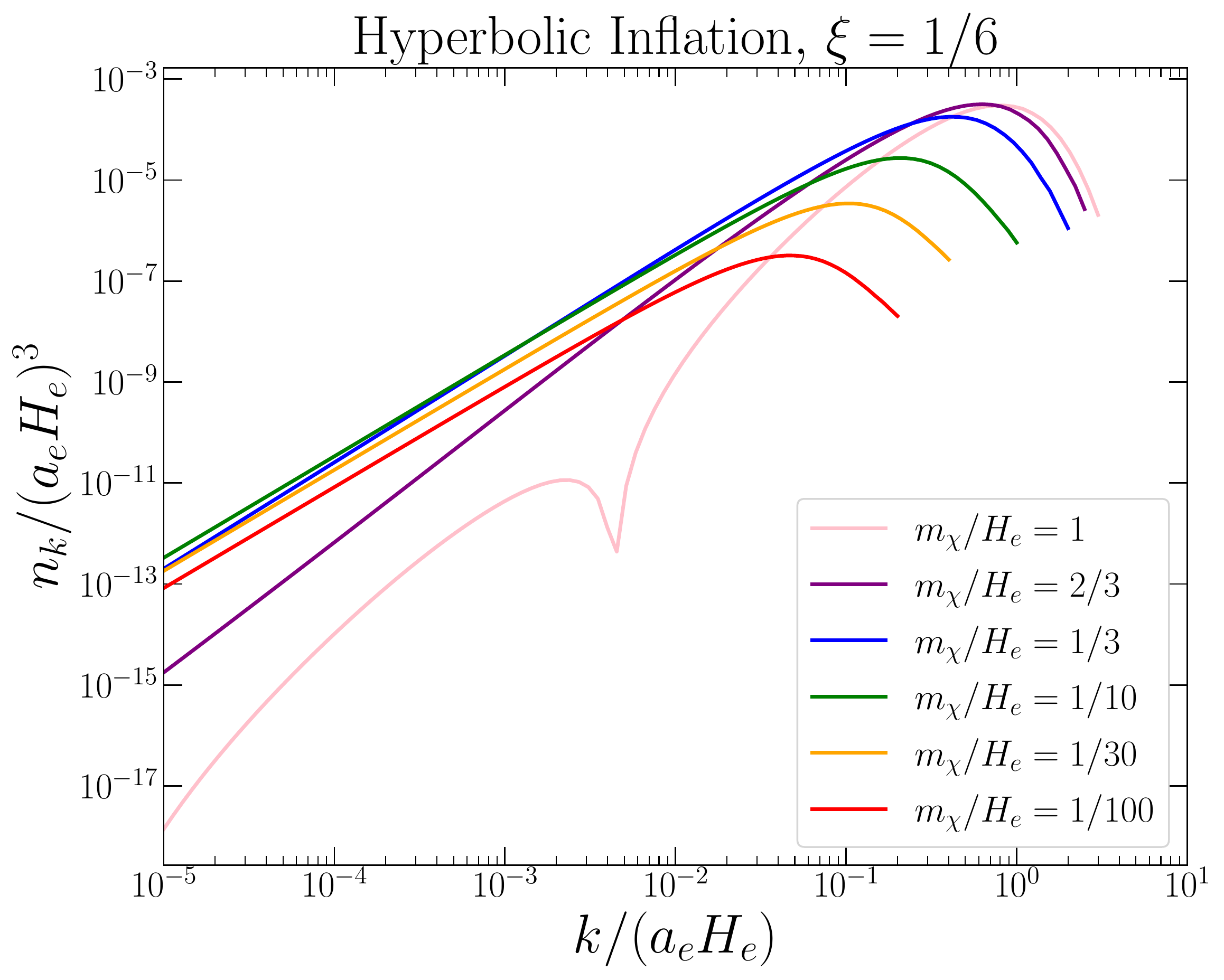}
	\includegraphics[width=0.49\textwidth]{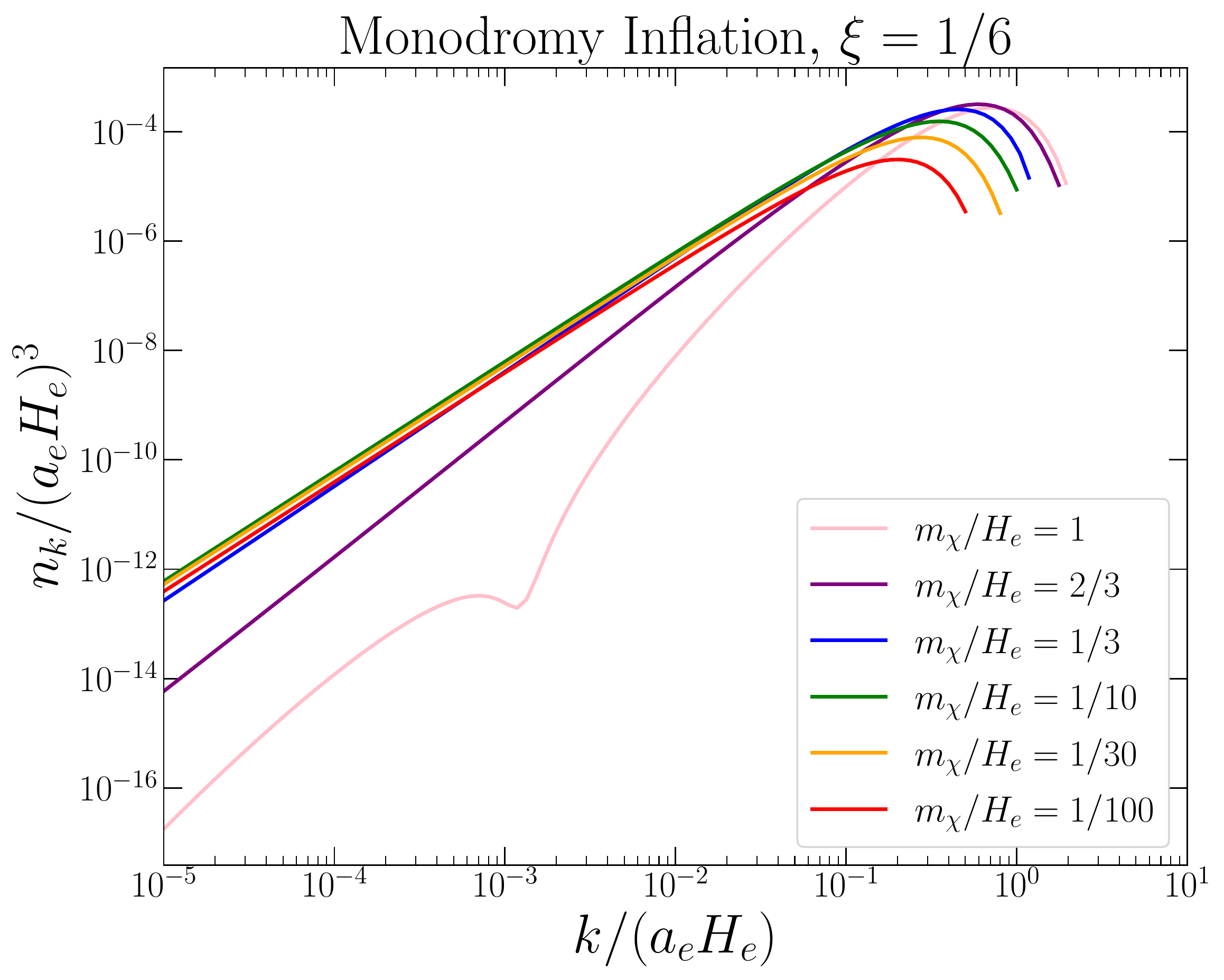}
	\caption{
\label{fig:n_k}
The spectrum of gravitationally produced spin-0 particles arising in models of rapid-turn multifield inflation. In each panel we show the late-time comoving number density spectrum $n_k$ as a function of the comoving wavenumber $k$.  Modes with $k / a_e H_e = 1$ are on the horizon scale at the end of inflation.  We study two models of rapid-turn multifield inflation: a model of hyperbolic inflation (left panels) and a model of monodromy inflation (right panels).  We study two models for the scalar spectator field:  a minimal coupling to gravity (top panels) and a conformal coupling to gravity (bottom panels).  Within each panel, the various curves correspond to different choices for the mass of the scalar spectator in terms of $H_e$, where $H_e$ is the Hubble parameter at the end of inflation ($H_e \sim \mathrm{GeV}$ in hyperbolic inflation and $H_e \sim 10^{12} \GeV$ in monodromy inflation).  } 
\end{figure}

Following the procedure described above, we numerically calculate the spectrum $n_k$ of gravitationally produced particles for the two models of rapid-turn multifield inflation and for several models of scalar spectator.  
These results are summarized in Fig.~\ref{fig:n_k}.  
Several notable features are worthy of further discussion.  

 {\bf Low-mass behavior:}  Comparing the two top panels of Fig.~\ref{fig:n_k} allows us to contrast the efficiency of gravitational particle production for low spectator masses, $\mu \equiv m_\chi / H_e \ll 1$, in the two models of inflation.  For hyperbolic inflation (top-left panel) the spectrum $n_k$ becomes insensitive to $m_\chi$ for $m_\chi \ll H_e$, whereas for monodromy inflation (right-panel) the spectrum grows like $(m_\chi/H_e)^{-1}$.  This difference can be traced to the post-inflationary dynamics: hyperbolic inflation is followed by a phase of kination ($w=1$) whereas monodromy inflation is followed by a phase of matter domination ($w=0$). As we show in App.~\ref{app:hyper}, the different redshifting factors account for the different scaling with $m_\chi$.  
 
{\bf IR behavior:}  The infrared (IR) modes ($k \ll a_e H_e$) exhibit a qualitatively different behavior as the spectator mass $m_\chi$ and gravitational coupling $\xi$ are varied.  For large masses $m_\chi \gtrsim H_e$ the spectrum is blue-tilted ($d\,\log\, n_k / d\,\log k > 0$), whereas for $m_\chi \ll H_e$ the spectral tilt is dictated by $\xi$.  A conformally-coupled scalar's spectrum ($\xi=1/6$, bottom panels) is blue-tilted, whereas a minimally-coupled scalar's spectrum ($\xi=0$, top panels) is nearly scale invariant and slightly red tilted -- a result which parallels the familiar calculation of inflaton fluctuations (see e.g., Ref.~\cite{Mukhanov:2005sc}).  A red-tilted spectrum can lead to an over-production of dark matter-photon isocurvature, which we discuss further below.  For the range of masses that we consider ($m_\chi / H_e > 10^{-2}$), we have numerically verified that $\xi \gtrsim 0.05$ is sufficient to yield a blue-tilted spectrum.  

{\bf UV behavior}: The ultraviolet (UV) modes ($k \gtrsim a_e H_e$) exhibit a suppression of gravitational particle production, as $n_k$ decreases toward high $k$ in each panel of Fig.~\ref{fig:n_k}.  This behavior can be understood analytically, since $\omega_k \approx k$ for large $k$, which corresponds to nearly adiabatic mode evolution with $A_k \to 0$ as $k \to \infty$.  

{\bf Model dependence}:  At minimal coupling, $\xi=0$, the number of produced particles in monodromy inflation is orders of magnitude larger than in hyperbolic inflation, while at conformal coupling $\xi = 1/6$ the number density of particles is comparable.  In the dispersion relation \eqref{eq:omegaksq} the Ricci-dependent term drops out for $\xi = 1/6$, which indicates that the dramatic difference between the two models at minimal coupling $\xi=0$ is due to the different behavior of the Ricci scalar after inflation: $R \propto -1/a^3$ following monodromy inflation and $R \propto -1/a^6$ following hyperbolic inflation.  

Some aspects of our numerical results can also be understood from simple analytical estimates, and we provide a detailed analysis in Appendix~\ref{app:hyper}. The strategy is to break up the evolution of a mode function into several time intervals where approximate analytical solutions are available.  Then the late-time solution is obtained by imposing continuity of the approximate solutions at the boundary of the intervals.  This approach has been utilized in e.g., Refs.~\cite{Ahmed:2019mjo,Kolb:2020fwh} where it was shown to capture the salient features. 

To highlight the power of this analytical analysis, we provide here an example where we seek to show why $n_k$ becomes insensitive to $m_\chi$ for hyperbolic inflation with a light scalar spectator $m_\chi < H_e$ that is minimally coupled to gravity $\xi = 0$.  For the remainder of this section we use the notation $\alpha = a/a_e$, $\mu = m_\chi/H_e$, and we set $a_e H_e = 1$.  We focus on modes with comoving wavenumber $k < 1$ that have left the horizon during inflation.  Modes that were initialized in the Bunch-Davies vacuum state inside the horizon, evolve towards a scale invariant spectrum outside the horizon, and at the end of inflation the mode function becomes 
\begin{equation}
	\lim_{\eta\rightarrow 0} |\chi_k|^2 \ \propto \frac{1}{k^3}
	\;.
\end{equation}
The subsequent evolution, after inflation, is qualitatively different for our two models of rapid-turn multifield inflation.  Here we focus on hyperbolic inflation, for which the post-inflationary dynamics have $w=1$, and the dispersion relation is given by 
\begin{equation}
	\omega_k^2 = k^2 + \frac{1}{\alpha^4} + \alpha^2 \mu^2 
	\;,
\end{equation}
which controls the mode function evolution through $\partial_\eta^2 \chi_k = - \omega_k^2 \chi_k$.   Consider the range of $k$ specified by $\mu^{4/3} < k < 1$.  For these modes, the mode function evolution can be broken up into three time intervals, corresponding to which of the terms in $\omega_k^2$ is dominant.   At first, the $k^2$ term dominates, and the modes are roughly constant in time $|\chi_k| \propto \alpha^0$.   At a threshold time $\alpha = \sqrt{k}$ the Ricci scalar term ($\alpha^2 R \propto 1/\alpha^4$) comes to dominate and the mode functions grow as $|\chi_k| \propto \alpha$.  Finally, the mass term becomes relevant at $\alpha = k/\mu$, at which point the modes begin to decay as $|\chi _k| \sim 1/\sqrt{\alpha}$.  Enforcing continuity of the mode function across the boundary between these time intervals leads to a solution during the final phase given by,
\begin{equation}
	 |\chi_k|^2 \propto \frac{1}{k^3 \mu \alpha} 
	 \;.
\end{equation}
From this expression one may compute late-time (but still within the $w=+1$ phase) particle number as, 
\begin{equation}
	\lim _{\rm late-time} n_k \approx \omega_k |\chi_k|^2  \simeq \frac{1}{2 \pi^2} \,\, \, \,\, (\mathrm{hyperbolic \,\,inflation})
\end{equation}
where we have used $\omega_k \rightarrow \alpha^2 \mu^2$ at late times. 
One may appreciate that the dependence on both the comoving wavenumber $k$ and the spectator mass $m_\chi = \mu H_e$ have completely cancelled. 
This is in agreement with the top-left panel of Fig.~\ref{fig:n_k}, which shows a nearly scale invariant spectrum for low-mass models $\mu = m_\chi / H_e \ll 1$. 
By contrast, a model of inflation that has $w=0$ for the post-inflationary dynamics, such as quadratic or monodromy inflation, would lead instead to $n_k \propto 1/\mu$ for the IR modes $k \ll a_e H_e$ and $\mu<1$.

\section{Dark Matter and Observable Constraints}
\label{sec:dm}

We now relate the GPP spectra, Fig.~\ref{fig:n_k}, to the present relic abundance. 
We will then proceed to map out the parameter space of the model that produces the observed dark matter abundance.

\subsection{Relic abundance}

We define the relic abundance of $\chi$ particles as $\Omega_\chi = \rho_{\chi,0} / \rho_c$ where $\rho_{\chi,0}$ is the energy density of $\chi$ particles today, $\rho_c = 3 \Mpl^2 H_0^2$ is the cosmological critical density, and $H_0 = 100 h \ \mathrm{km}/\mathrm{sec}/\mathrm{Mpc}$ is the Hubble constant.  For the model parameters that we study, the $\chi$ particles become nonrelativistic long before today, which lets us approximate $\rho_{\chi,0} \approx m_\chi n_{\chi,0}$ with $n_{\chi,0}$ their present-day number density.  If the comoving number density is conserved, which is the case if the $\chi$ particles only interact gravitationally, then we have $n_{\chi,0} a_0^3 = n_\chi a^3$, which can be evaluated using Eq.~\eqref{eq:nk}.\footnote{There is another potential contribution to $n_\chi$ from thermal freeze in~\cite{Garny:2015sjg,Sloth:2018wio,Garny:2017kha,Garny:2018grs}, e.g., reactions involving standard-model initial states,  like $\mathrm{SM} + \mathrm{SM} \to \chi + \chi$ via an $s$-channel graviton mediator.  This contribution depends on the mass of the dark matter particle and the reheat temperature, and in the parameter regime we consider is always subdominant to gravitational production.  Post-inflation production via the inflaton condensate, e.g., inflaton $+$ inflaton $\to \chi+\chi$ \cite{Tang:2017hvq,Mambrini:2021zpp} is automatically accounted for in the Bogouliov approach taken here \cite{Kaneta:2022gug}.   }  We calculate the relic abundance assuming that the spectator field becomes non-relativistic before the universe becomes radiation-dominated at reheating, i.e., the so-called `late reheating' scenario Ref.~\cite{Kolb:2020fwh}, and the details of this calculation appear in Appendix~\ref{app:relic abundance}.  Models of rapid-turn multifield inflation are distinguished by the equation of state during reheating, which is $w = 1$ for hyperbolic inflation and $w = 0$ for monodromy inflation.  For each of these models we find the relations (see Appendix \ref{app:relic abundance})
\begin{align}
\label{eq:relic-abundance-hyper}
	\frac{\Omega_\chi h^2}{0.12} 
	& \approx 
	\bigl( 2.5 \times 10^{-25} \bigr) 
	\biggl( \frac{m_\chi}{H_e} \biggr) 
	\biggl( \frac{T_\RH}{10^9 \GeV} \biggr)^{-1} 
	\biggl( \frac{H_e}{\GeV} \biggr)^3 
	\biggl( \frac{n_\chi a^3/a_e^3H_e^3}{10^{-5}} \biggr) 
	& (\textrm{hyperbolic inflation}) \\ 
\label{eq:relic-abundance-flat}
	 \frac{\Omega_\chi h^2}{0.12} 
	 & \approx 
	 \biggl( \frac{m_\chi}{H_e} \biggr) 
	 \biggl( \frac{T_\RH}{10^9 \GeV} \biggr) 
	 \biggl( \frac{H_e}{10^{12} \GeV} \biggr)^2 
	 \biggl( \frac{n_\chi a^3/a_e^3H_e^3}{10^{-5}} \biggr) 
	 &  (\textrm{monodromy inflation}) 
	 \;.
\end{align}
We have normalized $\Omega_\chi h^2$ to the measured dark matter relic abundance $\Omega_\DM h^2 \approx 0.12$.  The relation in Eq.~\eqref{eq:relic-abundance-flat} is familiar from earlier studies of gravitational particle production with a $w=0$ post-inflation phase. 

It is illuminating to compare Eqs.~\eqref{eq:relic-abundance-hyper}~and~\eqref{eq:relic-abundance-flat}.  For the fiducial parameters shown here, the monodromy model of inflation yields a relic abundance that is comparable to the observed dark matter relic abundance.  For hyperbolic inflation, the inflationary Hubble scale is much lower $H_e = \mathcal{O}(\mathrm{GeV})$, and for typical parameters we find that hyperbolic inflation significantly under-produces the dark matter.

\subsection{Isocurvature}
\label{sec:isocurvature}

Gravitational particle production leads to an approximately homogeneous distribution of $\chi$ energy density, but the inherently quantum nature of this production mechanism also generates inhomogeneities that do not align in space with the curvature perturbations generated by the inflationary sector.  If $\chi$ plays the role of dark matter, then this misalignment corresponds to a dark matter-photon isocurvature.  
Measurements of the cosmic microwave background radiation are consistent with the absence of isocurvature, and these observations impose constraints on an additional isocurvature component.  These constraints are expressed as an upper limit on the dimensionless isocurvature parameter $\beta_\mathrm{iso} = A_\Scal / A_\Rcal$ where $A_\Scal$ and $A_\Rcal$ are the amplitudes of the comoving isocurvature and curvature perturbation dimensionless power spectra at the CMB pivot scale.  Measurements by the \textit{Planck} satellite impose $\beta_{\rm iso}^{\rm CDM} < 0.038$ at the $95\%$ CL~\cite{Planck:2018jri}, assuming that isocurvature and curvature perturbations are uncorrelated.  In the models that we study, isocurvature is second-order in the field perturbations $\Scal \propto \delta \rho_\chi \sim m^2 (\delta \chi)^2$, and the the isocurvature power spectrum $\langle \Scal \Scal \rangle$ involves integrals over four $\chi$ mode functions~\cite{Ling:2021zlj}, which can be challenging to compute.  

In lieu of calculating the isocurvature power spectrum, we apply the following two diagnostic tests to assess the compatibility of our models with CMB data.  First, we calculate the $\chi$ relic abundance and impose $\Omega_\chi h^2 \leq 0.12$ to ensure that $\chi$ does not over-produce dark matter.  Second, we calculate the spectral tilt $n_\Scal -1 \equiv \mathrm{d} \log n_k / \mathrm{d} \log k$ of the low-$k$ power law and impose $n_\Scal - 1 \geq 0.2$.  If the spectrum $n_k$ is sufficiently blue-tilted, then the associated isocurvature power spectrum $\langle \Scal \Scal \rangle$ is also blue-tilted, and CMB constraints on dark matter-photon isocurvature are evaded; see Appendix~\ref{app:isocurvature} for additional details.  Since $\Omega_\chi h^2$ is primarily sourced by particles modes with $k \sim a_e H_e$ (see Fig.~\ref{fig:n_k}), while CMB isocurvature constraints are primarily sensitive to $k _{\rm CMB} \sim e^{-50} a_e H_e$, any small amount of blue-tilt is sufficient to render the amplitude of isocurvature on CMB scales well below the CMB constraint.  We impose $n_\Scal - 1 \geq 0.2$, which ensures that the amplitude of isocurvature at the CMB scale is suppressed by a factor of $e^{-10}$ relative to its value at $k \sim a_e H_e$, which is a priori suppressed by virtue of being second-order in perturbations ($\Scal \propto (\delta \chi)^2$).

\subsection{Quantitative results}
\label{subsec:results}

{\bf  Hyperbolic Inflation:}

In hyperbolic inflation, we find $\Omega_\chi h^2 \ll 0.12$ for all values of $m_\chi$ and $\xi$, and thus the model cannot match the observed dark matter abundance through gravitational production. The small relic density may be easily deduced from Eq.~\eqref{eq:relic-abundance-hyper}, and the fact that $H_e \sim {\rm GeV}$ and thus $T_\RH$ is at most $\mathcal{O}(10^{9}\mathrm{GeV})$, in hyperbolic inflation.  Therefore, it is always the case that too little dark matter is produced to match observations. We reiterate that the low scale of inflation is necessary to match the Planck constraint on the amplitude $A_s$ of the primordial power spectrum, despite the exponential growth of perturbations \cite{Mizuno:2017idt}. Thus this underproduction of dark matter is an unavoidable consequence of hyperbolic inflation satisfying CMB constraints.

One might wonder if particle production in hyperbolic inflation could be amplified in some region of parameter space, enough to compensate the low scale of inflation and match the observed abundance. However, we find numerically and show analytically in App.~\ref{app:hyper} that $n_k$ converges to a fixed form for small $m_\chi$ and $\xi=0$, limiting the number of particles that can be produced. This prevents the particle production from being amplified enough to compensate the small scale of inflation.

\noindent {\bf Monodromy Inflation:}

\begin{figure}[p]
\centering
\includegraphics[width=.7\textwidth]{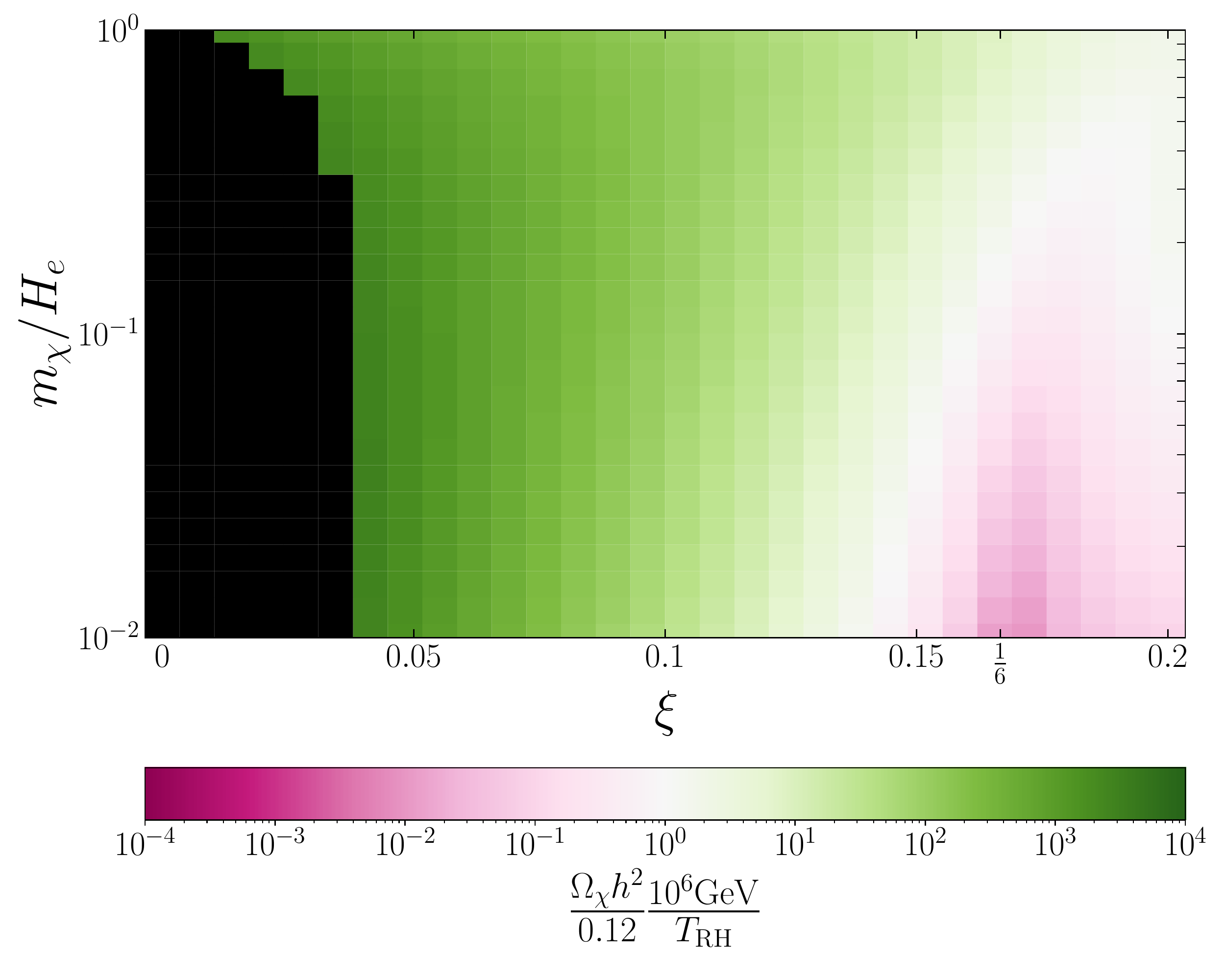} 
\includegraphics[width=.7\textwidth]{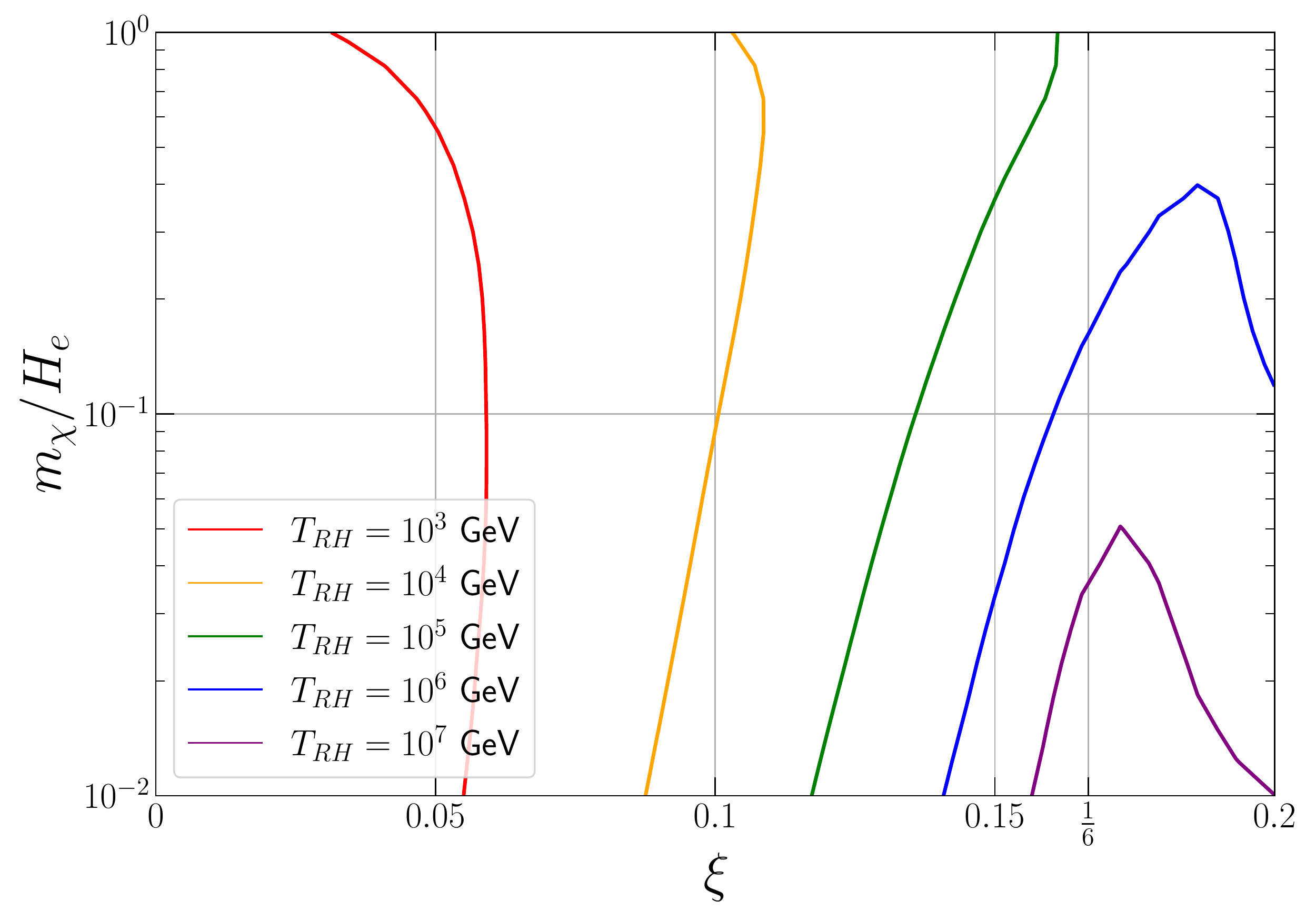}
\caption{
\label{fig:grid omega}
Predicted relic abundance over the parameter space.  The gravitational production of a scalar field $\chi$ with mass $m_\chi$ and non-minimal coupling to gravity $\xi$ is studied, and the present-day relic abundance $\Omega_\chi h^2$ is calculated as a function of these parameters.  Cosmological inflation is described by a model of rapid-turn monodromy inflation such that $H_e \simeq 1.4 \times 10^{13} \GeV$ at the end of inflation and the plasma is reheated to a temperature of $T_\RH$.  \textit{Top:}  For $T_\RH = 10^6 \GeV$ the predicted relic abundance $\Omega_\chi h^2$ is shown.  Along the white band, $\chi$ production saturates the observed dark matter relic abundance, $\Omega_\chi h^2 \approx \Omega_\DM h^2 \simeq 0.12$, while the green and pink regions correspond to over-production and under-production, respectively.  \textit{Bottom:} Values of $\xi$ and $\mu$ allowed by the relic abundance constraints, in monodromy inflation, for selected values of $T_\RH$.  }
\end{figure}

In the monodromy inflation model, the inflationary Hubble scale is $H_e =\mathcal{O}(10^{12}\mathrm{GeV})$, and it is possible for enough dark matter to be produced. In this case, $\Omega_\chi h^2 =0.12$ acts as a constraint on $m_\chi$, $\xi$ and $T_\RH$. In Fig.~\ref{fig:grid omega} we present the relic abundance of dark matter for varying $m_\chi$ and $\xi$, expressed as a function of $(\Omega_\chi h^2/0.12) (10^6 \GeV/T_\RH)$, using Eq.~\eqref{eq:relic-abundance-flat}. We focus solely on the region that satisfies Planck constraints on isocurvature (see Sec \ref{sec:isocurvature}), and we do not calculate the relic abundance of dark matter in the region defined by $n_\Scal -1 \leq 0.2$; this region appears in Fig. \ref{fig:grid omega} as empty black pixels. To further illuminate these results, in the bottom panel of Fig.~\ref{fig:grid omega} we also display contours of constant $T_\RH$ that give the observed dark matter abundance, $\Omega_\DM h^2 = 0.12$, i.e., curves of $(\Omega_\chi h^2/0.12) (10^6 \GeV/T_\RH) = 10^{-3}, 10^{-2}, 10^{-1}$, $1$, and $10$, for $T_\RH=10^9 \GeV$, $10^8 \GeV$, $10^7 \GeV$, $10^6 \GeV$,  $10^5 \GeV$, respectively. 

From Fig.~\ref{fig:grid omega} one may appreciate that the observed abundance of dark matter can be realized for a wide range of values of the coupling $\xi$ and reheat temperature $T_\RH$. Of particular note is that at minimal coupling ($\xi=0$), the observed DM abundance is realized for $m_\chi = {\cal O}(H_e)$ but only for $T_\RH = \mathcal{O}(10^3 \GeV)$. Thus, minimal coupling requires an extended period of matter domination, or rather, a delayed onset of radiation domination. Meanwhile, at conformal coupling ($\xi=1/6$) the observed DM abundance may be realized for $T_\RH \gtrsim 10^5 \GeV $ and mass $m < H_e$.

To contextualize these results, in Appendix  \ref{app:relic quadratic} we perform an identical analysis of GPP in $m^2\phi^2$ inflation, see Fig.~\ref{fig:relic_abundance_quadratic}. To allow for an `apples-to-apples' comparison between the two models, we adjust the parameter $m$ to provide an $H_e$ that matches that in the rapid-turn multifield inflation model.  Comparing Fig.~\ref{fig:relic_abundance_quadratic}  with Fig.~\ref{fig:grid omega}, one may see that  at any point $(\xi,m_\chi/H_e)$ the reheating temperature differs from the multifield case by a factor of less than 10. Additionally the two plots show almost identical qualitative behavior. This is a non-trivial check given the marked difference in the Ricci scalar $R$ in the two models; see Fig.~\ref{fig:comparison_alpha_R} in Appendix~\ref{app:relic quadratic}.

\begin{figure}[t]
\centering
\includegraphics[width=.7\textwidth]{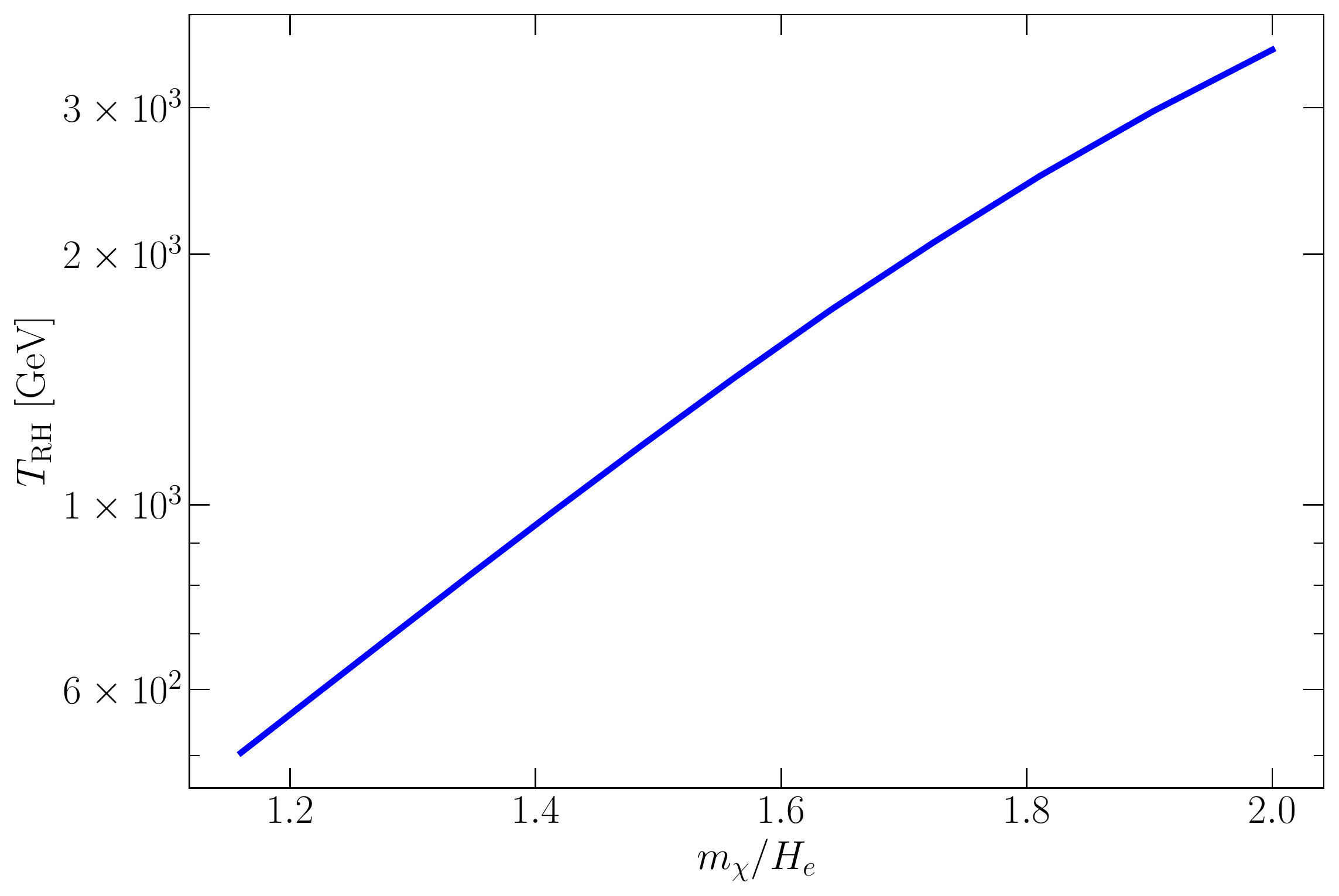}
\caption{Constraints on the super-Hubble-mass parameter space imposed by $\Omega_\chi \leq \Omega_\DM$.  Gravitational production of a minimally-coupled ($\xi=0$) scalar spectator field $\chi$ with mass $m_\chi$ leads to a relic abundance $\Omega_\chi h^2$ of non-relativistic particles that must not be larger than the observed dark matter relic abundance $\Omega_\DM h^2 \simeq 0.12$.  For a model of rapid-turn monodromy inflation with the Hubble expansion at the end of inflation taken to be $H_e \simeq 1.4 \times 10^{13} \GeV$, we obtain an upper limit on the post-inflationary reheating temperature $T_\RH$. }
\label{fig:Trh minimal}
\end{figure}

We extend our analysis to super-Hubble masses, $m_\chi/H_e >1 $, in Fig.~\ref{fig:Trh minimal}. We consider only minimal coupling ($\xi=0$) and a small range of $m_\chi/H_e$ due to the significant computational expense of each computation of $\Omega_\chi h^2$. From Fig.~\ref{fig:Trh minimal} one may appreciate the weak dependence of $T_\RH$ on $m_\chi$ for $m_\chi/H_e > 1$. We leave an investigation of the regime $m_\chi/H_e \gg1 $ to future work.

\section{Discussion}\label{sec:discussion}

In this work we have studied the gravitational production of spin-0 particles during the epoch of cosmological inflation and reheating by employing a recently-developed class of models, so-called rapid-turn multifield inflation.  The predictions of these models for the amplitude and spectrum of primordial curvature perturbations are consistent with cosmological observations, such as measurements of the CMB furnished by the Planck satellite telescope.  Additionally, models of rapid-turn inflation are theoretically compelling, because they naturally satisfy the de Sitter swampland conjecture, which would naively forbid slow-roll inflation.  We have demonstrated that the phenomenon of gravitational particle production in the context of rapid-turn monodromy inflation may be responsible for the origin of dark matter if the dark matter is made up of massive spin-0 particles that couple only to gravity (either minimally or for a range of non-minimal couplings).  

We have not attempted to perform a comprehensive scan of parameter space, either of the rapid-turn inflation models themselves or of the dark-matter model parameters (the dark-matter mass $m_\chi$ and the nonminimal coupling $\xi$). Instead, we have taken representative rapid-turn inflation models and fiducial inflation model parameters that yield observables consistent with Planck, and studied the feasibility of gravitational production of dark matter in these models. We find that rapid-turn inflation in monodromy inflation can produce the observed DM abundance for a wide range of masses and couplings, while rapid-turn inflation in a hyperbolic field generally can not. These analyses demonstrate the compatibility of rapid-turn multifield inflation with gravitational production of dark matter, and by extension, the compatibility of the dS swampland conjecture with the same.

There are many directions left for future work, such as an exploration of the model and parameter space of rapid-turn inflation models, and of other dark matter candidates, such as spin-1/2, spin-2, spin-3/2, or higher-spin. Also of interest is an exploration of the CMB non-Gaussianity generated by dark matter production in these models. We leave these and other exciting topics to future work. 

\section*{Acknowledgements}

The authors thank Vikas Aragam and Sonia Paban for insightful comments. The work of E.W.K. was supported in part by the US Department of Energy contract DE-FG02-13ER41958.  A.J.L. is supported in part by the National Science Foundation under Award No.~PHY-2114024.  EM and GP are supported in part by a Discovery Grant from the National Science and Engineering Research Council of Canada.

\begin{appendices}

\section{Analytic Estimate of GPP in Hyperbolic Inflation}
\label{app:hyper}

In the following are analytic approximations of $\displaystyle\lim _{\eta\to\infty} n_k$ as a function of $k$ and $\mu$, for monodromy inflation. We use the toy model presented in Table~\ref{tab:toy models w}. For simplicity, in this appendix we set $a_eH_e=1$ and we restrict to $k<1$ in what follows, i.e., modes that exit the horizon by the end of inflation. 

\begin{table}[t]
\begin{center}
\setlength{\tabcolsep}{20pt}
\renewcommand{\arraystretch}{1.5}\begin{tabular}{|r|c|c|c|} \hline
& $w$ & $\alpha(\eta)$ & $R(\eta)/6H_e^2$ \\ \hline
dS ($\eta<0$) & $-1$ & $(1-\eta)^{-1}$ & $-2$ \\ 
hyperbolic ($\eta>0$) & $+1$ & $\sqrt{1+2\eta}$ & $(1+2\eta)^{-3}$ \\ 
flat ($\eta>0$) & $\phantom{-}0$ & $\left(1+\dfrac{1}{2}\eta\right)^2$ & $-\dfrac{1}{2} \left(1+\dfrac{1}{2}\eta\right)^{-6}$ \\ \hline
\end{tabular}
\caption{  Toy model for the equation of state parameter $w$, the scale factor $\alpha=a/a_e$, and the Ricci scalar $R$ as a function of conformal time $\eta$ for hyperbolic inflation and monodromy inflation. We assume instantaneous and late reheating. }
\label{tab:toy models w}
\end{center}
\end{table}

\begin{figure}[t]
\centering
\includegraphics[width=0.75\textwidth]{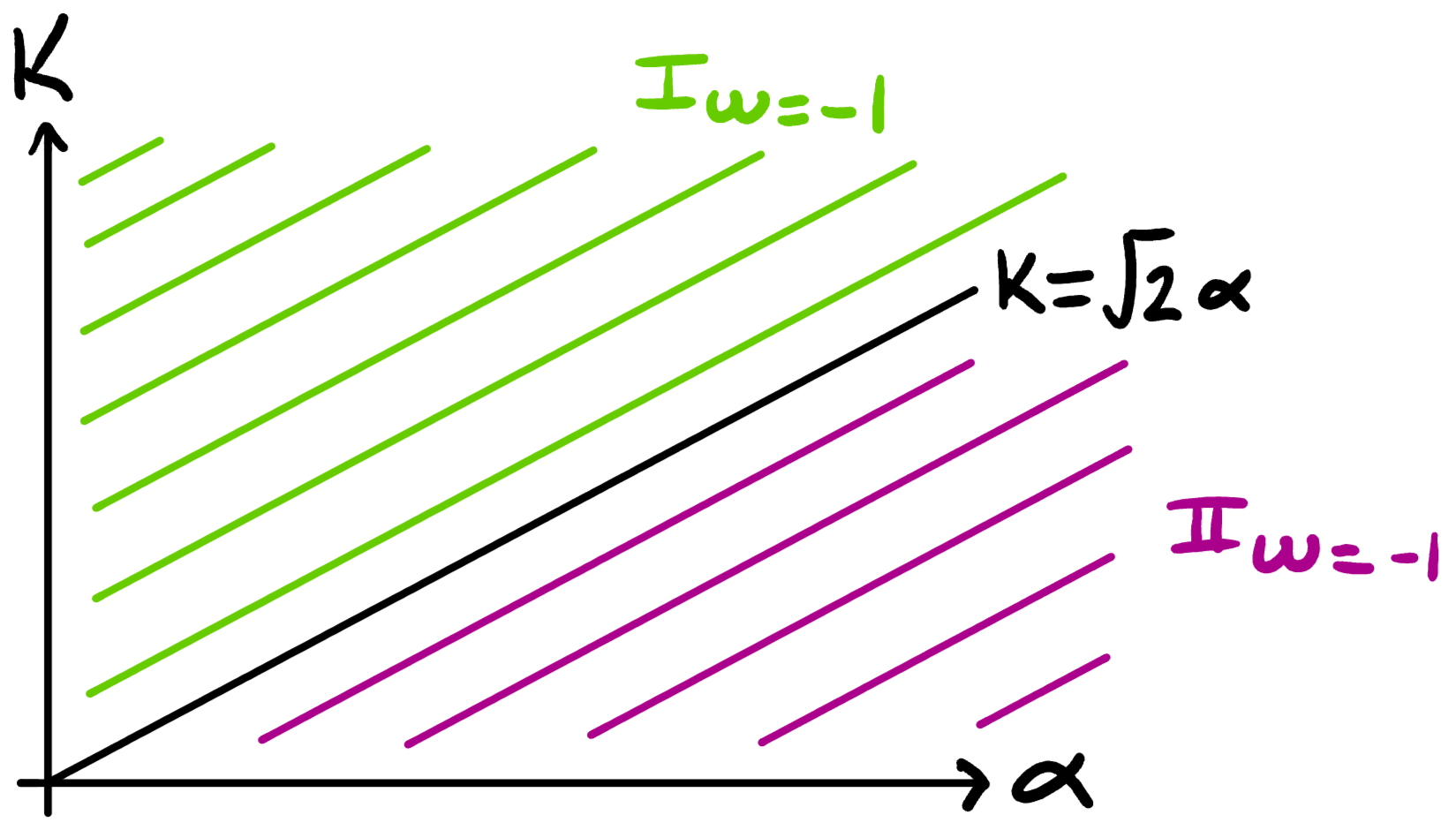}
\caption{Phase diagram in $\alpha$ and $k$ for $w=-1$. The modes only go through $\RN{1}_{w=-1}$ and $\RN{2}_{w=-1}$.}. 
\label{fig:phase diagram w=-1}
\end{figure}

During the de Sitter phase $\omega_k$ in Eq. (\ref{eq:EOMchi}) is given by
\begin{align}
    \omega_k^2&=k^2+\alpha^2\mu^2+\alpha^2\frac{R}{6H_e^2} =k^2+\alpha^2(\mu^2-2)
    \label{eq:wk w=-1}
\end{align}
where $\alpha \equiv a/a_e, \mu \equiv m_\chi/H_e$. Assume that $\mu < 1$, in which case we need to consider the two regions $\RN{1}_{w=-1}$ and $\RN{2}_{w=-1}$ shown in Fig. \ref{fig:phase diagram w=-1}. 

In each region, one of the terms in Eq. (\ref{eq:wk w=-1}) dominates and we ignore the other term. In particular, \begin{align}
    \omega_k^2 \approx\begin{cases}
    k^2 & \alpha<k/\sqrt{2} \hspace{0.5cm}(\RN{1}_{w=-1})\\
    -2\alpha^2 & \alpha> k/\sqrt{2}\hspace{0.5cm} (\RN{2}_{w=-1}) \ .
    \end{cases}
\end{align}
The EOM at $\alpha \approx 0$ is then that of $\RN{1}_{w=-1}$
\begin{align}
     \chi_k''+k^2\chi_k=0 \quad
    \implies \quad \chi_k = c_1e^{ik\eta}+c_2e^{-ik\eta} \ .
\end{align}
Applying the Bunch-Davies initial condition as $\eta \to -\infty$ gives in $\RN{1}_{w=-1}$
\begin{align}
    \chi_k = \frac{1}{\sqrt{2k}}e^{ik\eta} \quad \implies \quad |\chi_k|^2 = \frac{1}{2k}  \propto \alpha^0 \ .
\end{align}
Now moving on to $\RN{2}_{w=-1}$. The EOM is
\begin{align}
    &\chi_k''-2\alpha^2\chi_k=\chi_k''-\frac{2}{(1-\eta)^2}\chi_k=0 \nonumber \\
    \implies &\chi_k=c_1(1-\eta)^2+\frac{c_2}{1-\eta} = c_1\alpha^{-2}+c_2\alpha \ .
\end{align}
Keeping only the growing mode, $\chi_k = c_2\alpha$, so during evolution through $\RN{2}_{w=-1}$, $|\chi_k|^2$ grows as $|\chi_k|^2 \propto \alpha^2$ from $\alpha=k/\sqrt{2}$ to $\alpha=1$.  Therefore, as $\eta\to0$ ($\alpha\to1$)
\begin{align}
    |\chi_k|^2_{\eta\to0} &= \frac{1}{2k}\Big(\frac{1}{k/\sqrt{2}}\Big)^2 = \frac{1}{k^3}\ .
    \label{eq:enddS}
\end{align}
   
\begin{figure}[t]
\centering
\includegraphics[width=0.9\textwidth]{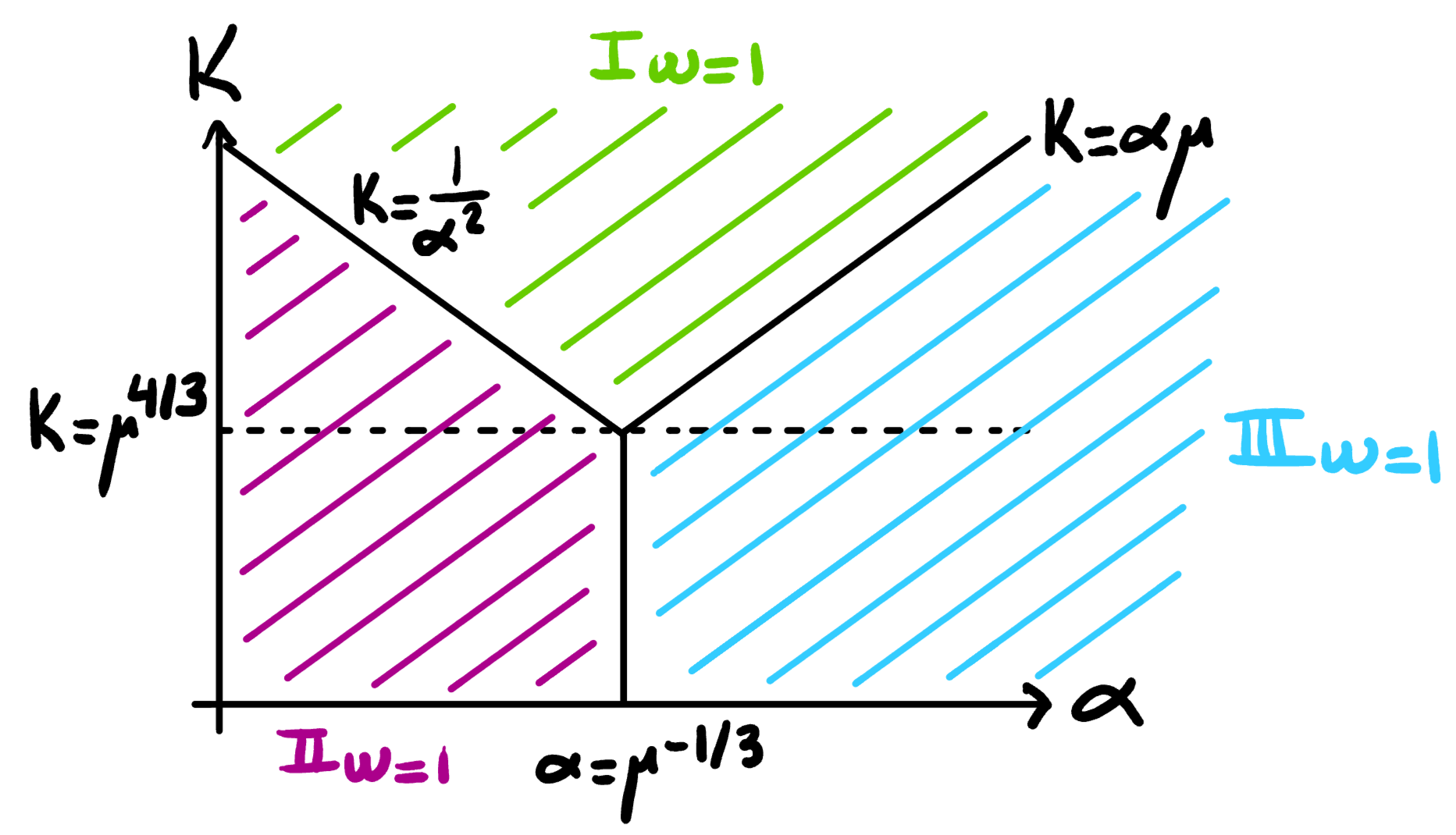}
\caption{Phase diagram in $\alpha$ and $k$ for $w=+1$. For $k>\mu^{4/3}$, the modes evolve in $\RN{2}_{w=1}$, then starting when $\alpha=1/\sqrt{k}$ the modes evolve through $\RN{1}_{w=1}$ until $\alpha=k/\mu$, then evolve in $\RN{3}_{w=1}$.  For $k<\mu^{4/3}$, the modes evolve in $\RN{2}_{w=1}$, then starting at $\alpha= \mu^{-1/3}$, they evolve in  $\RN{3}_{w=1}$. }\label{fig:phase diagram w=1}
\end{figure}

Now look at the $w=1$ phase. Using Table \ref{tab:toy models w}, $\omega_k$ in \ref{eq:EOMchi} is given by
\begin{align}
    \omega_k^2 = k^2 + \alpha^2\mu^2 + \frac{1}{\alpha^4}
    \label{eq:wk w=1}
\end{align}
Consider three regions $\RN{1}_{w=1}$, $\RN{2}_{w=1}$ and $\RN{3}_{w=1}$ shown in Fig. \ref{fig:phase diagram w=1}.

In each region, one of the terms in Eq. (\ref{eq:wk w=1}) dominates and we ignore the other terms. In particular, 
\begin{align}
    \omega_k^2 \approx\begin{cases}
    k^2 & k>1/\alpha^2 \ \ \& \   k>\alpha\mu  \hspace{1.0cm}(\RN{1}_{w=1})\\
    1/\alpha^4 & k<1/\alpha^2 \ \ \& \   \alpha<\mu^{-1/3}  \hspace{0.5cm} (\RN{2}_{w=1})\\
    \alpha^2\mu^2 & k<\alpha\mu \ \ \& \   \alpha>\mu^{-1/3}  \hspace{0.5cm}(\RN{3}_{w=1})
    \end{cases}
\end{align}
The EOM for $\RN{1}_{w=1}$ is $\chi_k''+k^2\chi_k=0$, with solution
\begin{align}
    &\chi_k = c_1e^{ik\eta} + c_2e^{-ik\eta} \implies |\chi_k|^2 \propto \alpha^0 \ .
\end{align}
The EOM for $\RN{2}_{w=1}$ is $\chi_k'' + \alpha^{-4}\chi_k = \chi_k''+(1+2\eta)^{-2}\chi_k = 0$, with solution
\begin{align}
  \chi_k = c_1\sqrt{1+2\eta} + c_2\sqrt{1+2\eta} \, \log(1+2\eta) = c_1 \alpha + c_2 \alpha \,\log(\alpha^2) \implies |\chi_k|^2 \propto \alpha^2 \ .
\end{align}
The EOM for $\RN{3}_{w=1}$ is $\chi_k'' + \alpha^2\mu^2\chi_k =  \chi_k'' + (1+2\eta)\mu^2\chi_k=0$, with solution
\begin{align}
\chi_k = c_1 \mathrm{Ai}\Big(-\Big(\frac{\mu}{2}\Big)^{2/3}\alpha^2\Big) + c_2 \mathrm{Bi}\Big(-\Big(\frac{\mu}{2}\Big)^{2/3}\alpha^2\Big) \implies  |\chi_k|^2 \propto \frac{1}{\alpha} \
\end{align}
where Ai and Bi are the Airy functions.   In the last step we used the fact that for large $x$, both $\mathrm{Ai}(-x^2)$ and $\mathrm{Bi}(-x^2)$ are proportional to $1/\sqrt{x}$.

Now we can evolve $|\chi_k|^2$ through de Sitter and kination to find the asymptotic value at late time. When finding $\displaystyle\lim _{\eta\to\infty} n_k$, there are two cases to consider: $k>\mu^{4/3}$ and $k<\mu^{4/3}$.  In both cases at the end of the de Sitter phase $|\chi_k|^2 = k^{-3}$ (recall Eq.~\eqref{eq:enddS}).

{\bf Case 1.} $k>\mu^{4/3}$:

In this case, starting at $\alpha=1$ with $|\chi_k|^2=k^{-3}$, the mode goes through $\RN{2}_{w=1}$ growing as $\alpha^2$ until $\alpha=1/\sqrt{k}$ when it enters $\RN{1}_{w=1}$ where it does not grow, then enters $\RN{3}_{w=1}$ when $\alpha=k/\mu$ and thereafter it decreases as $\alpha^{-1}$ (see Fig.~\ref{fig:phase diagram w=1}). The final result in $\RN{3}_{w=1}$ is
\begin{align}
|\chi_k|^2_{\alpha\to\infty} &= \frac{1}{k^3} \frac{1}{k}\frac{k/\mu}{\alpha} = \frac{1}{k^3\mu\alpha} \ .
\end{align}
Now, using that \eref{eq:bogoluibov} in the late-time limit $|\beta_k|^2_{\alpha\to\infty}= \omega_k|\chi_k|^2$,  and as $\alpha\to\infty$, $ \omega_k \to \alpha\mu$, we find
\begin{align}
n_k = \frac{k^3}{2\pi^2} |\beta_k|^2_{\alpha\to\infty} = \frac{k^3}{2\pi^2}\, \alpha\mu \, \frac{1}{k^3\mu\alpha} = \frac{1}{2\pi^2}
\end{align}

{\bf Case 2.} $k<\mu^{4/3}$:

In this case, again starting at $\alpha=1$ with $|\chi_k|^2=k^{-3}$, the mode goes through $\RN{2}_{w=1}$ growing as $\alpha^2$ until $\alpha=\mu^{-1/3}$ when it enters $\RN{3}_{w=1}$ where it evolves as $\alpha^{-1}$ (again, see Fig.~\ref{fig:phase diagram w=1}). 
The final result in $\RN{3}_{w=1}$ is
\begin{align}
|\chi_k|^2_{\alpha\to\infty} &= \frac{1}{k^3} (\mu^{-1/3})^2 \frac{\mu^{-1/3}}{\alpha} = \frac{1}{k^3\mu\alpha}\ ,
\end{align}
which is the same result as Case 1.

So in both cases $n_k\to (2\pi^2)^{-1}$, which has no dependence on $k$ or $\mu$. That $n_k$ asymptotes to a constant for small $\mu$ and $k$ can be readily observed in the numerics, see Fig.~\ref{fig:n_k}. This prevents the particle production from being amplified enough to compensate the low scale of inflation and match the observed abundance.

\section{Dark Matter Isocurvature}
\label{app:isocurvature}

Here we consider the isocurvature perturbations generated by gravitational production of dark matter. We relate the spectral index of isocurvature perturbations to the spectrum of produced particles $n_k$.

The isocurvature power spectrum is given as Ref.~\cite{Chung:2004nh},
\begin{equation}
    \Delta _\Scal^2 (q) = \frac{q^3}{2 \pi^2} \frac{1}{\bar{\rho}_\chi ^2} \frac{m_\chi^2}{a^6}\frac{1}{(2 \pi)^2} \displaystyle \int _{-1} ^1 {\rm d}z \int _{k_{\rm IR}} ^{k_{\rm UV}} {\rm d}k \, k^2 |\beta_k|^2 |\beta_{\sqrt{k^2 + q^2 - 2 k q z}}|^2 \ ,
\end{equation}
where $\bar{\rho}_\chi$ is the average mass density.   Consider the case that $\beta_k$ is peaked on length scale $k_*$ that is very large relative to the CMB pivot scale $k_{\rm piv}$, and assume that $\beta_k$ is monotonically decreasing as $k$ is made smaller, i.e., a blue-tilted spectrum with a constant spectral index.  Since $k^3 |\beta_k|^2$ drops off rapidly for $k>k_*$, we express the spectrum of $\beta_k$ for $k<k_*$ as 
\begin{equation}
    k^3 |\beta_k|^2  = A_\beta \left( \frac{k}{k_*} \right)^{n_\beta-1} \qquad (k<k_*) \ ,
\end{equation}
(a blue spectrum corresponds to $n_\beta>1$). In this case we may write the momentum integration as:
\begin{equation}
    \int  {\rm d}k \, k^2 |\beta_k|^2 |\beta_{\sqrt{k^2 + q^2 - 2 k q z}}|^2 = A_\beta^2 k_* ^{-2(n_\beta-1)}\int {\rm d} k\, k^{-2 + n_\beta} (k^2 + q^2 - 2kqz)^{(-4 + n_\beta)/2} \ .
\end{equation}

{\bf Case 1:} Consider the case that $n_{\beta} = 1 + \epsilon$ with $\epsilon < 1$. The integral is IR divergent and we may approximate it as,
\begin{equation}
     \int  {\rm d}k \, k^2 |\beta_k|^2 |\beta_{\sqrt{k^2 + q^2 - 2 k q z}}|^2 \sim A_\beta^2 k_*^{-2(n_\beta-1)}\, k_{\rm IR} ^{n_\beta-1} (k_{\rm IR}^2 + q^2 - 2k_{\rm IR} qz)^{(-4 + n_\beta)/2} \ . 
\end{equation}
Consider the case that $k_{\rm IR} \ll q \sim k_{\rm piv}$. In this case we have
\begin{equation}
    \int  {\rm d}k \, k^2 |\beta_k|^2 |\beta_{\sqrt{k^2 + q^2 - 2 k q z}}|^2 \sim A_\beta^2 k_* ^{-2(n_\beta-1)}\, k_{\rm IR} ^{n_\beta-1} q ^{-4 + n_\beta}  \ .
\end{equation}
We now can compute the spectral index of $\Delta_\Scal$ as
\begin{equation}
    n_\Scal - 1 \equiv  \frac{\rm d \log \Delta_\Scal(q)}{{\rm d} \log\,q } = \frac{d}{d \log\, q} \log \left( q^3 q ^{-4 + n_\beta} \right) = n_{\beta}-1
\end{equation}
Thus, for a slight blue spectrum of produced particles ($1< n_\beta <2$), the comoving isocurvature perturbation has the same spectral index as the produced particles, $n_\Scal = n_\beta$.

{\bf Case 2:} Consider the case that $n_\beta >2$. In this case the integral is dominated by the UV, and thus by the peak in the spectrum of produced particles at $k=k_*$. Since $q\ll k_*$, the integral is independent of $q$, and we find
\begin{equation}
    n_\Scal - 1 \equiv  \frac{\rm d \log \Delta_\Scal(q)}{{\rm d} \log\, q } = \frac{d}{d \log \, q} \log \left( q^3  \right) = 3
\end{equation}
Thus for $n_\beta >2$, the isocurvature spectrum has spectral index $n_\Scal-1=3$.

{\bf Punchline:} the comoving isocurvature perturbation $\Scal$ is blue-tilted provided that $k^3 \beta_k^2$ is blue-tilted.

\begin{figure}[t]
\centering
\includegraphics[width=0.9\textwidth]{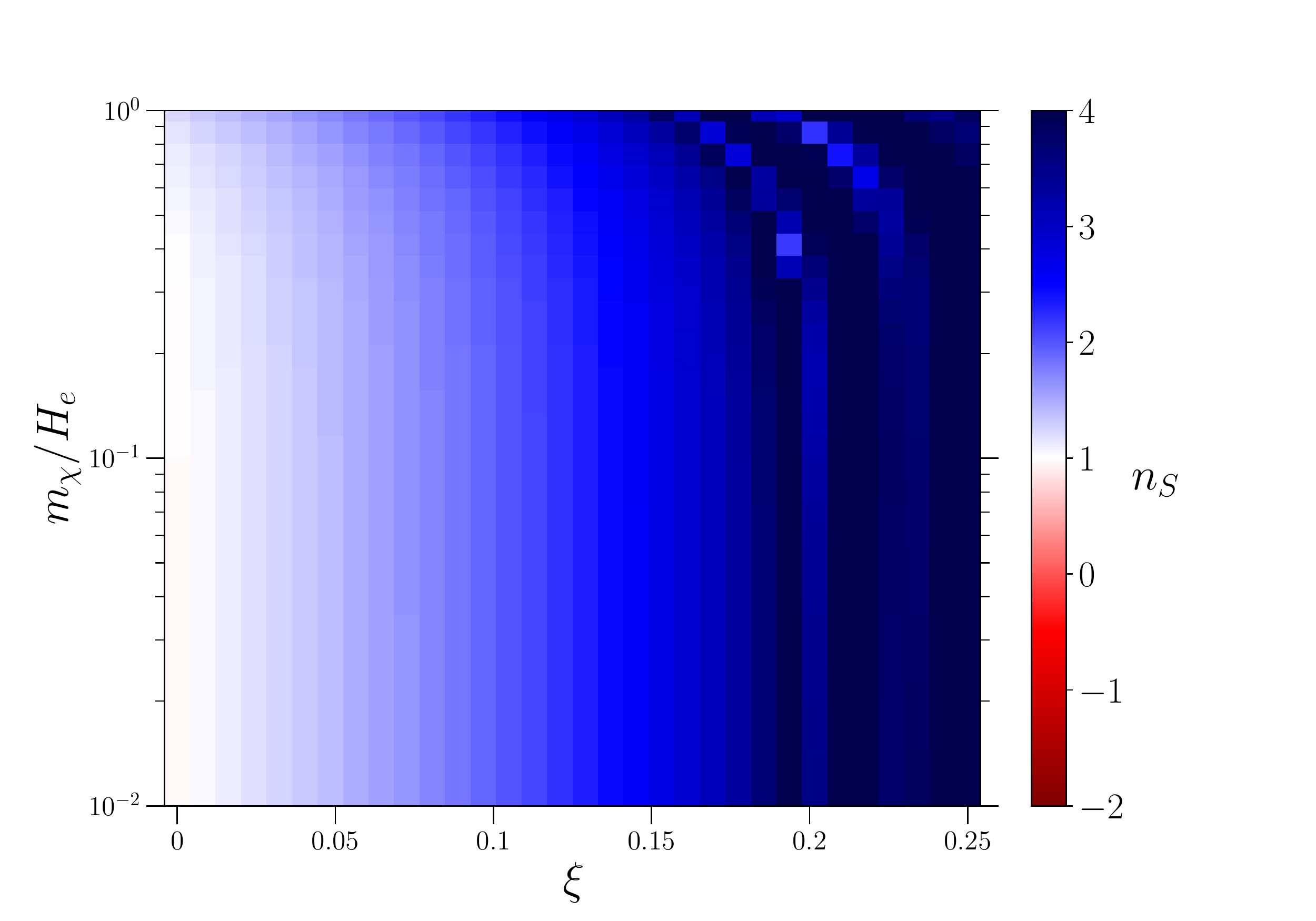}
\caption{Spectral index of isocurvature perturbations spectrum versus $\xi$ and $\mu$. The spectral index is calculated using the ratios of the values of $n_k$ at $k=10^{-4}$ and $k=10^{-5}$.}
\label{fig:spectral_tilt}
\end{figure}

\section{Relic abundance calculations }
\label{app:relic abundance}

The following is a derivation of Eqs.~\eqref{eq:relic-abundance-hyper} and \eqref{eq:relic-abundance-flat}. We wish to determine the relic abundance $\Omega$ of dark matter from the late time comoving number density $na^3$. We use the toy model for the equation of state parameter $w$ from the end of inflation to today. Inflation $(w=-1)$ is followed by a kination $(w=1)$ or matter-dominated ($w=0$) phase, then a transition to radiation domination $(w=1/3)$ at reheating, then a transition to matter domination $(w=0)$ at matter-radiation equality. Table \ref{tab:H dependence on a} presents the dependence of $H$ on $a$ for kination (K), radiation domination (RD), and matter domination (MD).

\begin{table}[!ht]
\begin{center}
\begin{tabular}{c|c|c|c|c}
& $w$ & $a$ & $\dot{a}$ & $H$\\
\hline
K & 1 & $t^{1/3}$ & $t^{-2/3}$ & $a^{-3}$\\
\hline
RD & 1/3 & $t^{1/2}$ & $t^{-1/2}$ & $a^{-2}$ \\
\hline
MD & 0 & $t^{2/3}$ & $t^{-1/3}$ & $a^{-3/2}$\\
\end{tabular}
\caption{Dependence of $H$ on the scale factor $a$ for kination (K), radiation domination (RD), and matter domination (MD). $a$, $\dot{a}$ and $H$ are given up to a multiplicative constant.}
\label{tab:H dependence on a}
\end{center}
\end{table}

We start with two expressions:
\begin{align}
\Omega_{\chi,0} & = \frac{\rho_{\chi,0}}{\rho_C} = \frac{m_\chi n_{\chi,0}}{3H_0^2\Mpl^2} \quad \mathrm{and} \quad
 n_{\chi,0}  = \frac{\left[n_\chi a^3/a_e^3H_e^3\right]}{a_0^3/a_e^3H_e^3} \ .
\end{align}
The factor of $\left[n_\chi a^3/a_e^3H_e^3\right]$ comes from the GPP calculation; we assume that the comoving number density $n_\chi a^3$ is subsequently conserved (i.e., additional production, decay, or annihilation processes are negligible).  
Now we outline the calculation of $a_e^3/a_0^3$.  Start by writing
\begin{align}
\frac{a_e^3}{a_0^3} = \frac{a_e^3}{\aRH^3}\times \frac{\aRH^3}{a_0^3}\ .
\end{align}
After reheating entropy is conserved, so up to factors of $g_S$, 
\begin{align}
\label{eq:as1}
 \frac{\aRH^3}{a_0^3} \sim \frac{T_0^3}{\TRH^3} \ .
 \end{align}
 The factor of $a_e/\aRH$ depends on whether the model is kination or matter dominated after inflation and before reheating.   From Table \ref{tab:H dependence on a}
 \begin{align}
 \label{eq:as2}
 \frac{a_e^3}{\aRH^3} = \left\{ \begin{array}{ll}
 \dfrac{\HRH}{H_e} & \mathrm{kination}\\[2ex]
\left( \dfrac{\HRH}{H_e}\right)^2 & \mathrm{MD}\ . \end{array} \right.
\end{align}
The final step is to relate $\HRH$ to $\TRH$ via $3\HRH^2\Mpl^2 = (\pi^2g_*/30)\TRH^4$.  Using Eqs.~\ref{eq:as1} and  \ref{eq:as2} along with the relationship between $\HRH$ and $\TRH$, and putting in the numerical factors results in Eqs.~\eqref{eq:relic-abundance-hyper} and \eqref{eq:relic-abundance-flat}.

\section{Dark matter production in quadratic inflation}
\label{app:relic quadratic}

\begin{figure}[t]
\centering
\includegraphics[width=.7\textwidth]{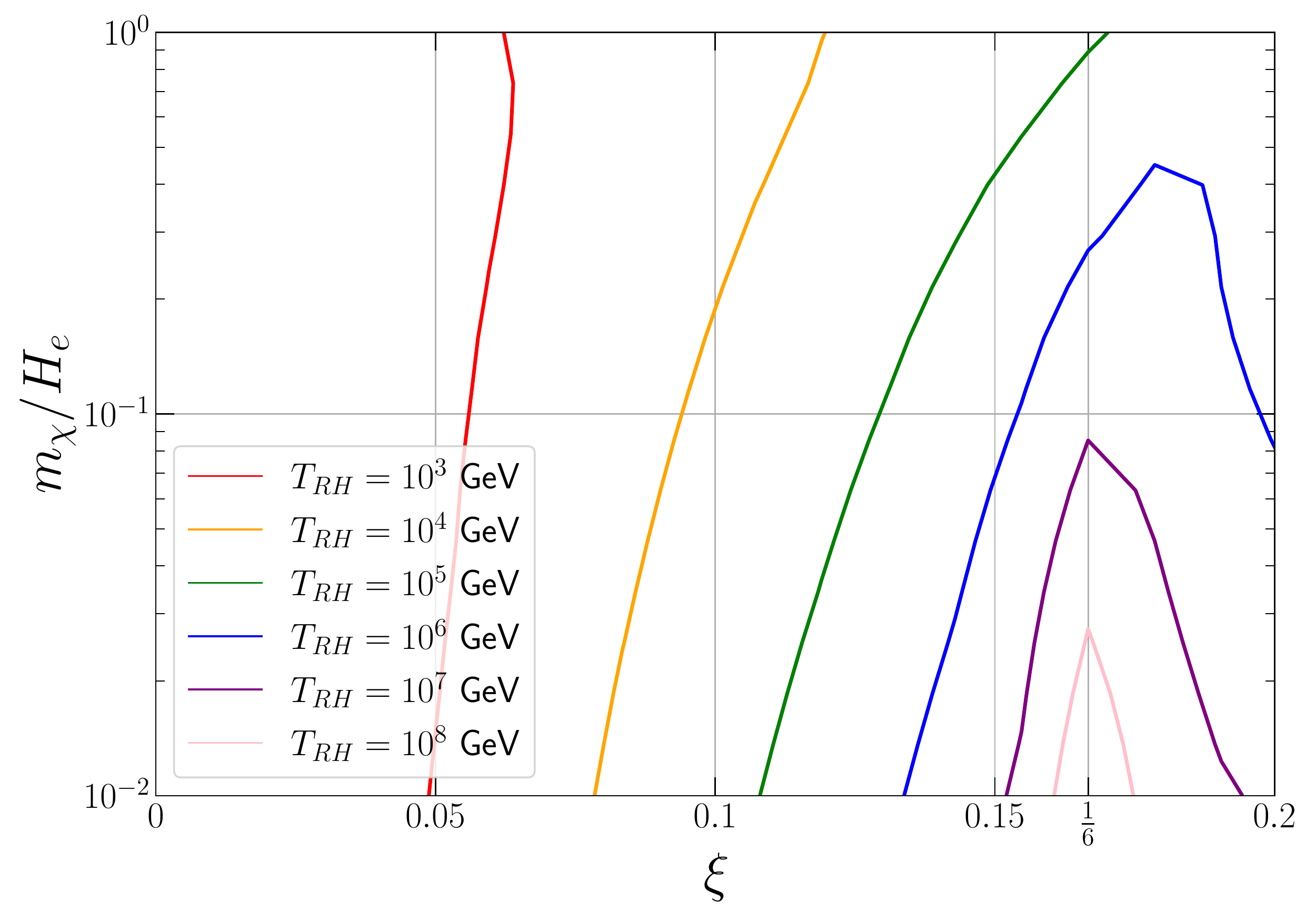}
\caption{Dark matter production in $m^2\phi^2$ inflation. Values of $\xi$ and $m_\chi/H_e$ allowed by the relic abundance constraints for selected values of $T_\RH$. Cf., Fig.~\ref{fig:grid omega}.}
\label{fig:relic_abundance_quadratic}
\end{figure}

Here we highlight the similarities between GPP of dark matter in the monodromy inflation model (see Sec. \ref{subsec:results}) and in $m^2\phi^2$ inflation with $H_e$ artificially rescaled to match that in the multifield model. Fig.~\ref{fig:relic_abundance_quadratic} presents a plot identical to the bottom panel of Fig.~\ref{fig:grid omega} except that it is computed in $m^2 \phi^2$ quadratic inflation with $m$ chosen so as to provide the desired $H_e = 5.8 \times 10^{-6} \Mpl \approx 1.4 \times 10^{13} \GeV$.

 At any point $(\xi,\mu)$ the reheating temperature differs from the multifield case by a factor of less than 10. Additionally the two plots show almost identical qualitative behavior. An observable difference is the behavior of the reheating temperature contours approaching $\mu=1$. For that reason we expect that the two plots would show more significant quantitative differences at $\mu >1$.

\begin{figure}[t]
\centering
\includegraphics[width=.7\textwidth]{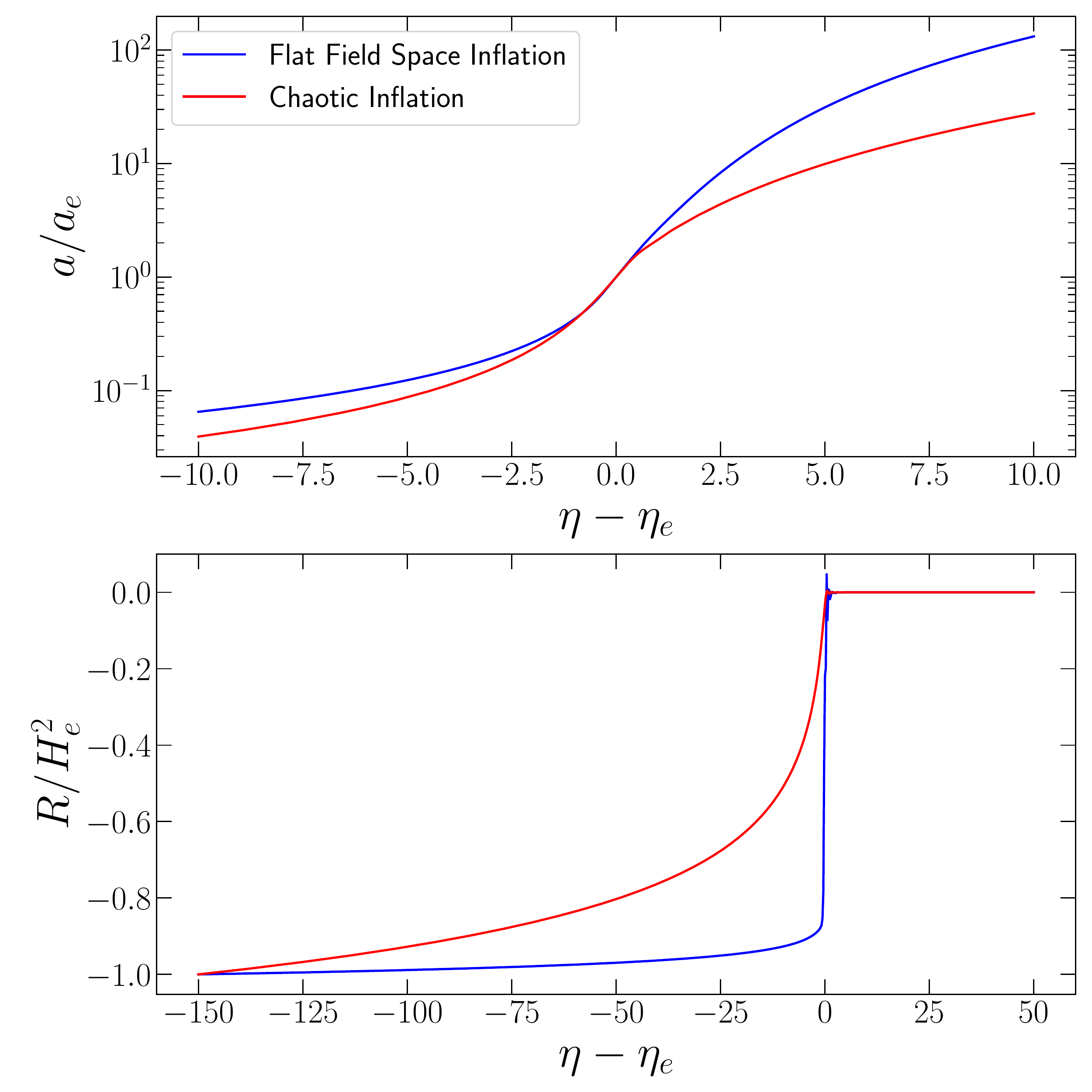}
\caption{Evolution of the scale factor $a$ and Ricci scalar $R$ in monodromy inflation versus $m^2\phi^2$ inflation with $H_e$ rescaled to match the multifield model. For better comparison, both Ricci scalars are normalized to $-1$ at $\eta=-150$. Both $a$ and $R$ are visibly different between the two models, highlighting that the similarities in the GPP of dark matter we observe between the two models were not known a priori.}
\label{fig:comparison_alpha_R}
\end{figure}

To highlight that the similarity in GPP of dark matter in monodromy inflation and in $m^2\phi^2$ inflation could not be known with certainty prior to our calculations, we compare in Fig. \ref{fig:comparison_alpha_R} the evolution of the scale factor and Ricci scalar in both inflation models. Since these two quantities govern through Eq. \ref{eq:omegaksq} the evolution of dark matter mode functions during inflation, their differences could have resulted in more severe disparities in GPP.

\end{appendices}

\bibliographystyle{JHEP}
\bibliography{refs}

\end{document}